\documentclass[aps,twocolumn,showpacs,superscriptaddress,prb,10pt]{revtex4-1}
\usepackage{graphicx}
\usepackage{caption}
\usepackage{dcolumn}
\usepackage{bm}
\usepackage{color}
\usepackage{amsmath}
\usepackage{amsfonts}
\usepackage{subfig}
\usepackage[makeroom]{cancel}

\newcommand{\COMMENTED}[1]{}
\newcommand{\REMARKS}[1]
{
{ \color{red}{\textbf{ {[#1]} }} }
}

\begin{document}

\author{Adam Chiciak}
\affiliation{Department of Physics, The College of William and Mary, Williamsburg, Virginia 23187}

\author{Ettore Vitali}
\affiliation{Department of Physics, The College of William and Mary, Williamsburg, Virginia 23187}

\author{Hao Shi}
\affiliation{Department of Physics, The College of William and Mary, Williamsburg, Virginia 23187}
\affiliation{Center for Computational Quantum Physics, Flatiron Institute, 162 5th Avenue, New York, New York 10010}

\author{Shiwei Zhang}
\affiliation{Department of Physics, The College of William and Mary, Williamsburg, Virginia 23187}




\title{Magnetic orders in the hole doped three-band Hubbard model: \\
spin spirals, nematicity, and ferromagnetic domain walls} 

\begin{abstract}
The Copper-Oxygen planes in cuprates have been at the center of the
search for a theory of high-temperature superconductivity.  We conduct
an extensive study of the ground state of the three-band Hubbard
(Emery) model in the underdoped regime. We focus on the magnetic and
charge orders,
and present results
from generalized Hartree-Fock (GHF) calculations.
The ground-state properties 
at the thermodynamic limit are challenging to pin down because of 
sensitivity to computational details including the shapes and sizes of the supercells.
We employ large-scale computations with various technical 
improvements to determine the orders within GHF.
\COMMENTED{
 are scanned to determine
the magnetic and charge properties.  We present mean field results
from generalized Hartree-Fock calculations on large lattice sizes,
aiming to capture the different kinds of orders that are stabilized in
the thermodynamic limit.
}  
The ground state exhibits a rich phase diagram 
with hole doping as the charge transfer energy is varied,
including ferromagnetic domain walls embedded in an antiferromagnetic background,
spin spirals, and nematic order.
\end{abstract}

\maketitle

\section{Introduction}
\label{sec:intro}

Despite more than thirty years of intense theoretical \cite{Fischer_nematic,Poilblanc_HF,Zaanen_Gunnarsson,Millis_stripes,PhysRevB.92.205112,White_Scalapino_stripes,MACHIDA1989192,Kato_SDW,Sarker_spiral,Thomson_Sachdev,White_checkerboard,RevModPhys.66.763} 
and experimental \cite{Tranduada_AIP,Tranquada_stripes,Fujita_stripes,Chang_nature,Comin_xray,Jurkutat_NMR,Rybicki2016,Achkar_Xray,Haase_NMR,RUAN20161826,Weber_chargetransfer} efforts,
understanding high temperature superconductors 
has remained a major challenge.
Significant progress has been made\cite{RevModPhys.78.17,RevModPhys.82.2421,RevModPhys.87.457}
(see, e.g. Refs 22-24 for some reviews).
The advent of 
petaflop computing and the emergence of new and improved computational methods
have 
created unprecedented opportunities for the computational study of these exciting systems.

It is widely believed that the physical mechanism underlying high temperature superconductivity of
the cuprates arises in the quasi--two--dimensional physics in the copper--oxide planes \cite{Emery},
with the other layers of the materials playing the role of charge reservoirs that 
can chemically dope the planes by adding or removing holes. Starting from a parent compound (zero doping) which is an antiferromagnetic insulator \cite{Armitage_Mott}, 
the antiferromagnetism rapidly melts as holes are added, giving rise to a rich phase diagram where
charge and spin orders appear to coexist, cooperating or competing \cite{RevModPhys.87.457, PhysRevLett.94.156404} with d-wave pairing superconducting order.
A crucial question, naturally, is the following: what is the best model that captures the most
important physics while being simple enough to allow 
theoretical and computational
approaches to provide accurate answers?

The 
majority of recent theoretical efforts 
to study 
these planes have focused on the Hubbard hamiltonian \cite{Anderson} which, in the realm of high temperature superconductivity, is a low energy effective model that relies 
on the Zhang--Rice singlet notion \cite{PhysRevB.37.3759}.
These studies assume that the explicit contribution of the oxygen degrees of freedom can
be neglected. Although impressively accurate results\cite{Mingpu-sc-PRB} have been obtained
on the one--band Hubbard model 
and very interesting magnetic and charge
orders have emerged 
which reproduce 
some important experimental results, there are indications that the model, at least at zero temperature, 
does not allow a superconducting phase to become stable in the experimentally relevant 
regions of the parameter space (The t--J model, although closely related, can contain different physics\cite{Sorella_tJ,Plakida_tJ,Bejas_tJ,Sachdev_tJ}
from the one--band Hubbard model with ``physical'' parameters for the interaction strength). 

Moreover, recent x-ray scattering and nuclear magnetic resonance experiments seem to indicate 
that 
spin and charge density waves involve the Oxygen $p$-bands in a 
non-trivial manner \cite{Comin_xray,Jurkutat_NMR,Rybicki2016,Achkar_Xray,Haase_NMR}.
This suggests that one important direction to build a model that captures the physics 
of the cuprates may be the implicit or explicit inclusion of the Oxygen orbitals.

With advances in computational resources, it is now viable for a variety of methods
to go beyond the one-band 
Hubbard model 
and treat the more realistic, three-band Hubbard
model, known as the Emery model \cite{Emery}, which explicitly includes the Copper $3d_{x^2-y^2}$
and the Oxygen $2p_x$ and $2p_y$ orbitals. 
The Hamiltonian contains a set of parameters, including charge transfer energy, hopping amplitudes and on-site repulsion
energies, which can in principle be computed with {\it{ab initio}} approaches, resulting in standard parameter sets for the cuprates. 

However, the actual choice of the parameters
is subtle. In particular, the value of the charge-transfer energy,
$\Delta\equiv \epsilon_p - \epsilon_d$, can be affected by double-counting issues \cite{Wang_double} in the computation.
Empirically, the value of $\Delta$ appears to vary across different families of cuprates, controlling the density of holes around Copper and Oxygen atoms, and there are indications from experiment that it 
is correlated, or more precisely anticorrelated,
with the critical temperature \cite{RUAN20161826,Jurkutat_NMR,Rybicki2016,Weber_chargetransfer}. 
Recently we have seen from quantum Monte Carlo calculations that the properties of the undoped, parent 
system can vary fundamentally with $\Delta$ \cite{Ettore_three-band}.

The prominent role of the parameter $\Delta$ motivated 
us to scan its values here and investigate the effects
on the physical properties. The role of the charge-transfer energy has been 
 investigated theoretically using cluster dynamical mean field theory \cite{0295-5075-100-3-37001}, 
where the pairing order was addressed.
Here we focus on spin and charge orders in the underdoped regime,
which play an integral role in understanding the origin of superconductivity and potentially how to 
optimize the critical temperature $T_c$.
\COMMENTED{
 whose long range behavior is not accessible to cluster approaches: we argue that the dependence of structural and magnetic properties on the charge-transfer energy, and thus
their relation to the critical temperature, can shed important light into the very important problem
of what is the best environment in which superconductivity can become stable.
}
Although several studies addressed the behavior of the Emery model using different methodologies,
including exact diagonalization of small clusters \cite{Dobry_ED}, 
random phase approximation \cite{Atkinson_RPA},
embedding methodologies \cite{Arrigoni_cluster, PhysRevB.78.035132}, 
quantum Monte Carlo \cite{Dopf_QMC, Scalettar_QMC,Yanagisawa_3bandQMC,CPMC_3band}
and Density Matrix Renormalization Group \cite{PhysRevB.92.205112}, 
comprehensive knowledge of the ground-state phase diagram, especially a full 
analysis of the role
of $\Delta$ 
in determining the properties of the model, is still missing.

Since the phase diagram  in the underdoped regime is intrinsically complex, with several cooperating and competing phases separated by small energy scales, it is crucial to be able to study very large lattices with different geometries and boundary conditions in order to rule out
spurious finite size effects. 
A useful illustration is found in the one-band 
Hubbard model, where
the magnetic and charge orders in the ground state exhibit long wavelength collective modes which are extremely delicate
and sensitive to finite-size effects \cite{chia-chen-prb-08,chia-chen-prl-10,science-17}.
The task to systematically determine such orders is especially challenging for many-body approaches, with
high computational costs that tend to make it particularly difficult to reliably reach the thermodynamic limit 
and scan multiple parameters.  
Mean-field calculations can serve as a valuable guide in this regard. In the one-band Hubbard model,
unrestricted Hartree-Fock solutions were found to capture much of the magnetic phase diagram
\cite{Xu-JPCM-11}, albeit with 
parameter values that can differ \cite{Mingpu-sc-PRB,science-17}, reflecting the tendency of mean-field theory to exaggerate order.

\COMMENTED{

we believe, 
requires 
a thorough study with mean-field 
mandatory to pave the road for more sophisticated many-body approaches. A direct study with a cutting-edge method, in our opinion, could miss the right physics due to a wrong lattice geometry or similar; mean-field is cheap enough to allow us to perform many experiments, in order to find a reasonable variational state. 
}

Although mean-field studies have been performed on the three-band Hubbard model\cite{Zaanen_Gunnarsson,Assaad-prb-93,Pan_3band,Seibold_3bandHF,Seibold_polarize,Sadori_stripes,Seibold_canting},
a systematic determination of 
the magnetic and charge properties within a general mean-field framework is,
to our knowledge, still not available. 
This is the goal of the present work. 
We employ a generalized Hartree-Fock (GHF) approach which allows non-colinear magnetic orders.
As described below,
we find a rich ground-state phase diagram,
with some phases hereforeto not seen in models for cuprates.
The phase diagram of the three--band model is much more complex than 
that of the single-band Hubbard model. 
Based on the lessons from the single-band model, these phases should serve as very plausible 
candidate  zero-temperature states of the model, possibly with modified parameters (e.g., reduced 
effective repulsion). 
The results identify several phases which are potentially important to the 
physics of the CuO$_2$ planes in cuprate materials.
Our results also
provide guidance on finite-size and other subtleties for many-body calculations.   
The mean-field solution can serve as possible trial wave functions for more advanced quantum Monte Carlo (QMC) investigations.


The rest of the paper is organized as follows. In Sec.~\ref{sec:model} we introduce the 
three-band Hubbard, or Emery,
model. In Sec.~\ref{sec:method}, 
we outline the GHF 
method and explain the need, and our strategies, for 
scanning the parameter space. 
Sec.~\ref{sec:results}
presents our resultsin which three distict phases are observed as $\Delta$ is varied: 
\ref{ssec:magDW}.~magnetic domain walls, 
\ref{ssec:spirals}.~spin spirals,  
and 
\ref{ssec:intermediate}.~nematic phases in the intermediate $\Delta$ region.
We further discuss the results and conclude in Sec.~\ref{sec:concl}.

\section{Model}
\label{sec:model}

\begin{figure}[ptb]
\includegraphics[width=6.0cm, angle=0]{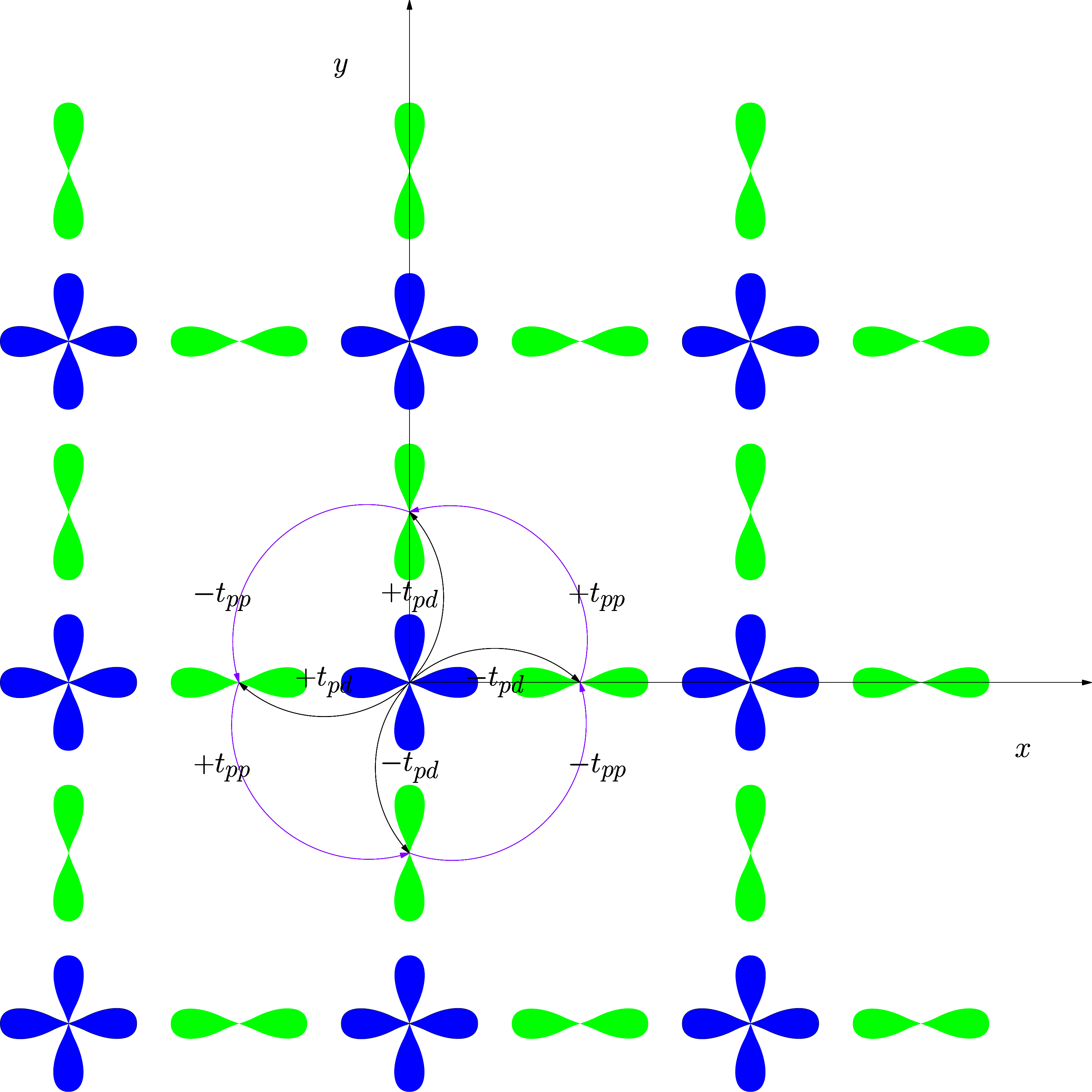}
\caption{
 (Color online) Schematic view of the CuO$_2$ 
 planes
 in 
 cuprates. 
 Cu $3d_{x^2 - y^2}$ orbitals are represented in blue,
 and $O$ $2p_x$ and $2p_y$ orbitals in green.
 We use the reference frame defined by the two axes in the figure.
 The curve connectors represent the hopping, and the labels
 define the sign rule.}
\label{fig:model}
\end{figure}

The Emery model includes the Cu $3d_{x^2 - y^2}$ orbital and 
the O $2p_x$ and $2p_y$  orbitals  in the description of the  
cuprates. 
In Fig.~\ref{fig:model}, 
a schematic representation of the CuO$_2$ plane is given to help  visualize the model.
The Hamiltonian is 
\begin{equation}
\label{3bands:ham}
\begin{split}
& \hat{H} = \varepsilon_d \sum_{i, \sigma} \hat{d}^{\dagger}_{i,\sigma} \hat{d}^{}_{i,\sigma} 
+ \varepsilon_p \sum_{j, \sigma} \hat{p}^{\dagger}_{j,\sigma} \hat{p}^{}_{j,\sigma} + \\
& \sum_{<i,j>, \sigma} t_{pd}^{ij} \left(\hat{d}^{\dagger}_{i,\sigma} \hat{p}^{}_{j,\sigma} + h.c \right)
+  \sum_{<j,k>, \sigma} t_{pp}^{jk} \left(\hat{p}^{\dagger}_{j,\sigma} \hat{p}^{}_{k,\sigma} + h.c \right)
\\
& + U_d  \sum_{i} \hat{d}^{\dagger}_{i,\uparrow} \hat{d}^{}_{i,\uparrow} \hat{d}^{\dagger}_{i,\downarrow} \hat{d}^{}_{i,\downarrow} + U_p \sum_{j} \hat{p}^{\dagger}_{j,\uparrow} \hat{p}^{}_{j,\uparrow} \hat{p}^{\dagger}_{j,\downarrow} \hat{p}^{}_{j,\downarrow}\,.
\end{split}
\end{equation}
In Eq.~\eqref{3bands:ham}, $i$ runs over the sites $\vec{r}_{\rm Cu}$ of a square lattice
$\mathbb{Z}^2$ defined by the positions of the Cu atoms.
The labels $j$
and $k$ run over the positions of the O atoms, shifted with respect
to the Cu sites, $\vec{r}_{{\rm O}_x} = \vec{r}_{\rm Cu} + 0.5 \, \hat{x}$ for the $2p_x$ orbitals, and
$\vec{r}_{{\rm O}_y} = \vec{r}_{\rm Cu} + 0.5 \, \hat{y}$ for the $2p_y$ orbitals. 
This model is formulated in terms of holes rather than electrons: for example, the operator $\hat{d}^{\dagger}_{i,\sigma}$
creates a hole on the $3d_{x^2 - y^2}$ orbital at site $i$ with spin $\sigma = \uparrow,\downarrow$.
The first two terms in the 
Hamiltonian contain the orbital energies, which define the charge-transfer
energy parameter $\Delta = \varepsilon_p - \varepsilon_d$,  the energy needed for a hole
to move from a Cu $3d_{x^2 - y^2}$ orbital to an O $p$ orbital.
The next two terms 
describe hopping between orbitals; the hopping amplitudes $t_{pd}^{ij}$ and $t_{pp}^{jk}$ are expressed in terms
of two parameters, 
$t_{pd}$ and $t_{pp}$,
and the dependence on the sites is simply a sign factor, 
as depicted in  Fig.~\ref{fig:model}.
Finally, the last two terms 
represent the on-site repulsion energies, namely
double-occupancy penalties similar to those in the one--band Hubbard model. We neglect Coulomb interactions beyond the on-site terms.

As mentioned in the introduction, 
we study the properties of the model as a function
of the charge transfer energy $\Delta$. 
Our starting point is an {\it{ab intio}} set \cite{Wagner_abinitio} of parameters obtained for La$_2$CuO$_4$, the parent compound of
the lanthanum based family of cuprates.
The parameter values are listed in Table~\ref{tab:param_table}.
\COMMENTED{
 $\varepsilon_p = -3.2$ eV, $\varepsilon_d = -7.6$ eV,
$t_{pd} = 1.2$ eV, $t_{pp} = 0.7$ eV, $U_p = 2$ eV and $U_d = 8.4$ eV.
}
This set corresponds to a charge transfer energy $\Delta = 4.4$ eV.
To correct for possible double counting issues \cite{Wang_double} would imply 
a considerable reduction of this value
to $\Delta \sim 1.5$ eV.
In our calculations we scan the value of $\Delta $ in this range.
As we will show below, the variation of $\Delta$ can dramatically affect 
the physical properties of the model.

\begin{table}
\begin{center}
  \begin{tabular}{ l c  c  c  c c c}
    \hline\hline
      Parameter & $U_d$ & $U_p$ & $\varepsilon_d$ & $\varepsilon_p$ & $t_{pd}$ & $t_{pp}$ \\
       \hline
	Value (eV) & 8.4 & 2.0 & -7.6 & -3.2 & 1.2 & 0.7\\
	\hline\hline
  \end{tabular}
\end{center}
\caption{Parameter values adopted in the present study. 
The parameters are obtained from  \cite{Wagner_abinitio} La$_2$CuO$_4$. 
We scan the value of 
$\Delta = \varepsilon_p - \varepsilon_d$ from $4.4$ to $1.5$ in our study.
}
\label{tab:param_table}
\end{table}

\section{Methods}
\label{sec:method}

In order to study the Hamiltonian in Eq.~\eqref{3bands:ham}, we use a fully self-consistent 
mean-field approach, where the two-body 
operators are decoupled as follows:
\begin{equation}
\label{3bands:mf}
\begin{split}
&  \hat{d}^{\dagger}_{i,\uparrow} \hat{d}^{}_{i,\uparrow} \hat{d}^{\dagger}_{i,\downarrow} \hat{d}^{}_{i,\downarrow} \simeq {\rm const}   \\
& + \langle  \hat{d}^{\dagger}_{i,\uparrow} \hat{d}^{}_{i,\uparrow} \rangle \, \hat{d}^{\dagger}_{i,\downarrow} \hat{d}^{}_{i,\downarrow} +  \langle  \hat{d}^{\dagger}_{i,\downarrow} \hat{d}^{}_{i,\downarrow} \rangle \, \hat{d}^{\dagger}_{i,\uparrow} \hat{d}^{}_{i,\uparrow} \\
& - \langle  \hat{d}^{\dagger}_{i,\uparrow} \hat{d}^{}_{i,\downarrow} \rangle \, \hat{d}^{\dagger}_{i,\downarrow} \hat{d}^{}_{i,\uparrow} -  \langle  \hat{d}^{\dagger}_{i,\downarrow} \hat{d}^{}_{i,\uparrow} \rangle \, \hat{d}^{\dagger}_{i,\uparrow} \hat{d}^{}_{i,\downarrow}\,. 
\end{split}
\end{equation}
The terms in brackets, densities and spin densities, are obtained self-consistently by a diagonalization of a one-body Hamiltonian. 
We use a generalized Hartree-Fock approach, which constrains only the total number of particles (holes), $N$.
%
\COMMENTED{
In our calculations we 
frequently introduce random noise in the orders and anneal the solution to help
 the mean-field procedure locate the order corresponding to the global minimum in energy.
}


Since we are interested in the bulk physics of the model, we perform calculations on
large lattices, Cu$_{L}$O$_{2L}$ hosting $N$  holes corresponding to a doping of $h = N/L - 1$.
In order to minimize size effects, which is important if 
long-range charge or magnetic orders are present,  we explore different choices of geometries (square and rectangular supercells) and boundary conditions while keeping $h$ fixed.
More precisely we consider both periodic boundary conditions (PBC) and 
twist boundary conditions (TBC) with random twist angles.
We examine both ``regular'' and tilted supercells.
The former has  basis vectors 
 along the coordinates axes depicted in Fig.~\ref{fig:model}, while the latter has basis vectors along the two diagonal directions, i.e.,
those obtained from rotating the $x$- and $y$-axes by 
$\pi/4$. 
The tilted supercell 
requires considering unit cells containing two Copper atoms and four Oxygen atoms and is meant to capture orders along the diagonal direction. 
In both cases, we will use the notation $L_a\times L_b=L$ to denote the dimension of 
our supercell, with $a$ ($b$) being either $x$ ($y$) or the two diagonal directions. We 
study systems with increasing $L$ to improve the extrapolation to the bulk limit.

For given lattice and number of holes, the GHF solution will be a Slater determinant of spin orbitals:
\begin{equation}
\begin{split}
& | \Psi \rangle = \prod_{n=1}^{N} \hat{\phi}^{\dagger}_n \, |0 \rangle \\
& \hat{\phi}^{\dagger}_n = \sum_{i=1}^{L} \sum_{\sigma=\uparrow,\downarrow} \sum_{\alpha=d,p_x,p_y} u_n(i,\alpha,\sigma)
\, \hat{\alpha}^{\dagger}_{i,\sigma}
\end{split}
\end{equation}
that minimizes the energy $\langle \Psi \, | \, \hat{H} \, | \, \Psi \rangle$ within the manifold of
Slater determinants. From the wave function $|\Psi\rangle$, any ground-state 
property of the model can be computed. 
The GHF approach allows the number of the spin-$\uparrow$ and spin-$\downarrow$ particles
to fluctuate and 
non-colinear spin orders to develop.
No particular order is assumed at the beginning of the calculation. 
The self-consistent procedure can lead to a local minimum.
%
In our calculations we 
frequently introduce random noise in the orders and anneal the solution to help
 the mean-field procedure locate the order corresponding to the global minimum in energy.

\section{Results}
\label{sec:results}

\COMMENTED{
This is a comprehensive study of the mean field states of the copper oxide plane. 
The parameter space is more complicated than that of the one--band model. We consider the physical parameters
given by the table below.
However, the ab initio calculations for the charge transfer energy, 
$\Delta = \varepsilon_p - \varepsilon_d$, may be plagued by double counting issues.
Since the charge transfer energy is important in determining the d and p-pand occupation in the 
three--band model, it is important that we study the system while varying the charge transfer energy.}

\begin{figure}
\includegraphics[width=8.0cm, angle=0]{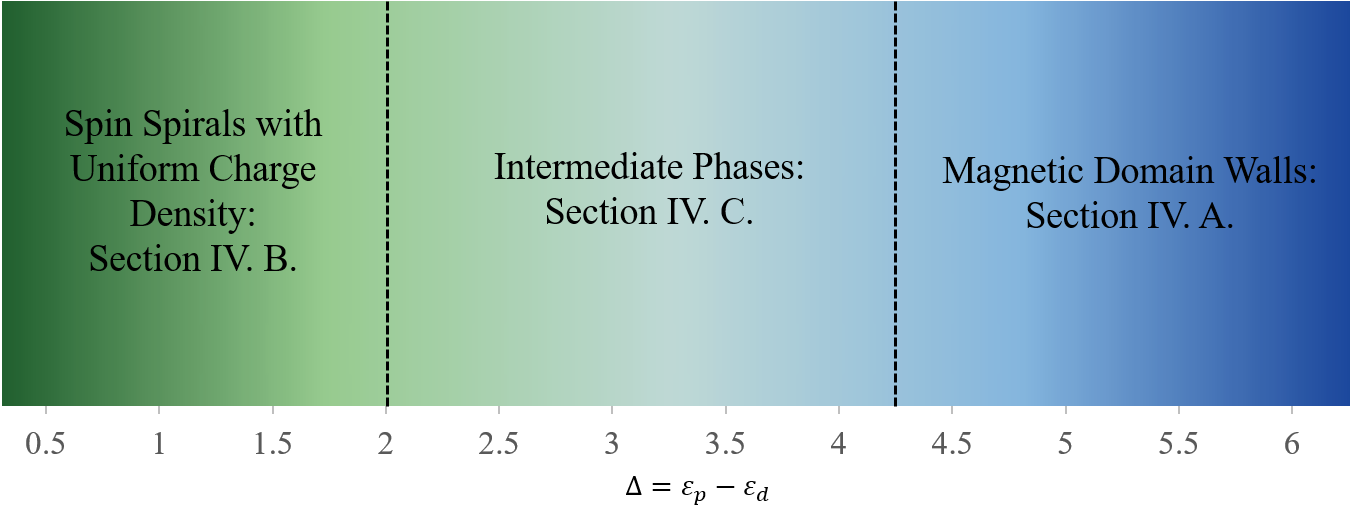}
\caption{
(Color online) Different phases given by GHF as a function of the charge transfer energy $\Delta$.
The phase ``boundaries'' are meant as rough guides only.
Each parameter regime is discussed in a subsection as indicated in the figure.
}
\label{fig:phase_table}
\end{figure}

Most of our calculations are at doping $h = 1/8$, using the hopping 
and on-site interaction parameters given in Table~\ref{tab:param_table}, while 
scanning the charge-transfer energy $\Delta$,
%
which represents
the energy needed to move a hole from a Copper atom to an Oxygen atom.
The parameter  $\Delta$, via its
interplay with hopping amplitudes and on-site Coulomb repulsion, 
has a direct consequence on 
the average density of holes at the Cu and O sites: $n_d$ and $n_p$, respectively. 
Magnetic resonance experiments \cite{Rybicki2016} are now able to detect the values of $n_d$ and $n_p$  for the different families of cuprates.
These results indicate
that smaller differences $(n_d - n_p)$, i.e. smaller values of $\Delta$, correspond to higher critical temperature and thus higher superfluid fraction. 
The anticorrelation between $\Delta$ and the critical temperature is also seen 
from scanning tunneling microscopy \cite{RUAN20161826}. 
Cluster dynamical mean field theory
calculations indicate  \cite{0295-5075-100-3-37001} similar tendencies.
Thus a systematic investigation of the dependence of the ground-state properties on $\Delta$ is 
especially important and timely.

\COMMENTED{
Even if, within our mean field approach, we cannot have access to superfluidity, we can explore
bulk charge and spin order as a function of $n_d - n_p$, or equivalently of $\Delta$. We argue that the evolution of those orders as $n_d - n_p$ is decreased will put light
into the crucial question about which is the most favorable scenario for the superconducting order to become stable.
}

We find a rich and complex set of possible ground-state magnetic and charge orders as $\Delta$ is scanned. 
At hole doping of $h = 1/8$ and with hopping and on-site repulsion parameters listed in 
Table~\ref{tab:param_table},  three different regimes are encountered with GHF as $\Delta$ is varied:
magnetic domain walls at 
high $\Delta$ ($\sim 4.4$\,eV), spiral spin-density waves 
at low $\Delta$ ($\sim 1.5$\,eV), and nematic  intermediate phases in between.
The gradient plot in Fig.~\ref{fig:phase_table}
shows the parameter range scanned, and these different regimes, which are described separately in 
the following three subsections.


\COMMENTED{
\begin{center}
        \begin{tabular}{r|l|l|l|l|l|l}
        Parameters & $U_d$ & $U_p$ & $\varepsilon_d$ & $\varepsilon_p$ & $t_{pd}$ & $t_{pp}$ \\
        \hline
	Value (eV) & 8.4 & 2.0 & -7.6 & -3.2 & 1.2 & 0.7
        \end{tabular}
	\label{fig:param_table}
\end{center}
}

\subsection{Magnetic Domain Walls (MDWs)}
\label{ssec:magDW}

\subsubsection{Spin and Charge Order at Doping h = 1/8}
\label{subbssec:magDW-1o8}

\begin{figure}[ptb]
\begin{center}
\includegraphics[width=7.0cm, angle=0]{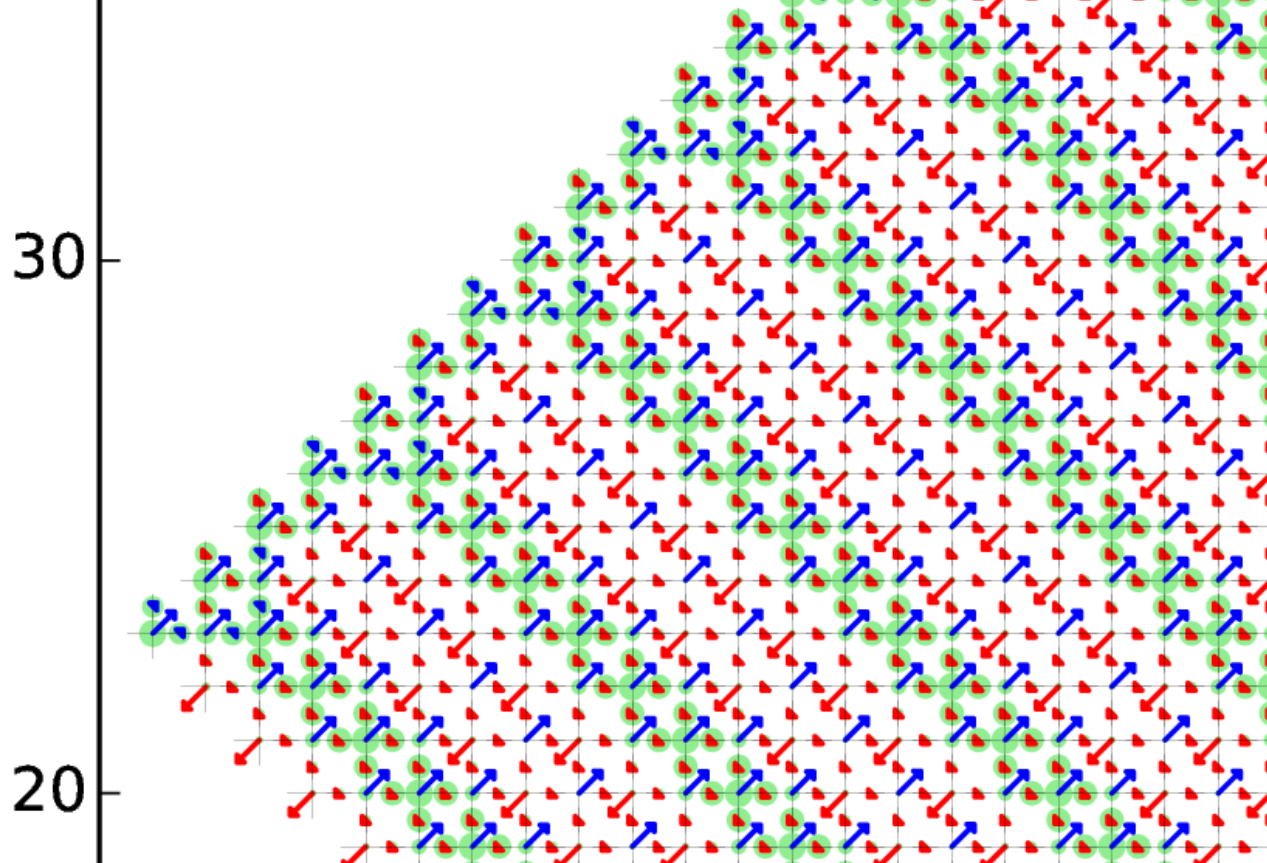}
\caption{
(Color online) Enlarged section of the magnetic and charge order 
in the ground state of a $24\times 30$ lattice at $\Delta = 4.4\,eV$, $h = 1/8$,
and tilted (diagonal) PBC. 
All spins are aligned or anti-aligned along one  (arbitrary) direction in this state.
The spin is plotted as red (positive)  and blue (negative) arrows,
with their length representing the magnitude.
The excess hole density given by $\delta n_{\rm excess}$ is proportional to the size of the green circles.
}
\label{fig:24x30}
\end{center}
\end{figure}

We start from the largest value of the charge-transfer energy, $\Delta = 4.4$ $eV$.
At half-filling, the system is antiferromagnetic (AFM) and the densities of holes on 
the Copper and Oxygen sites are 
$n_d \approx 0.735$ and $n_p \approx 0.133$, respectively.
At doping of $h = 1/8$, the system exhibits diagonal magnetic
domain walls (MDW) in the ground state.
In Fig.~\ref{fig:24x30} we show the spin and charge densities, using a large supercell, 
$24 \times 30$, with tilted 
periodic boundary conditions, i.e., Cu$_{1440}$O$_{2880}$, in order to determine the order.
We find that the spins
($S_x(\vec{r}) , S_y(\vec{r}), S_z(\vec{r})$) in the GHF ground state are, to very good accuracy, aligned (anti-aligned) in one arbitrary direction. In the plot we use arrows to represent the spin density, 
and circles to represent 
 the excess hole density, 
which we define as:
\begin{equation}
\delta n_{\rm excess}(\vec{r}) = n_{\rm doped}(\vec{r}) - n_0(\vec{r})\,,
\end{equation}
where $n_0(\vec{r})$ is the density at 
half-filling.

From the plot in  Fig.~\ref{fig:24x30},
we clearly see an array of lines of increased hole density, with $n_d \approx 0.842$, i.e., 
extra occupation on the order of $\delta n_{\rm excess} \sim 0.1$,  forming a $\pi/4$ angle with respect
to the CuO bonds, superimposed to the AFM 
background. 
On the Cu $d$-sites within these domain wall lines, there is a near perfect spin flip ($\vec{S} \rightarrow -\vec{S}$)  creating local 
ferromagnetic order.
These $d$-sites are surrounded by four O $p$-sites which also have increased
hole occupancy,  $n_p \approx 0.225$. 
 The doped holes are concentrated on these lines, with only slight
``spill-over'' to the adjacent AFM lines. These MDW structures are spaced out and
 embedded in the
 rest of the system which is kept essentially at the AFM
state found at half-filling.
The spin on the O $p$-sites, negligible at half--filling, remains very small but does show noticible increase (${\mathcal O}(10^{-3})$).
The small spins on the $p$-sites are all aligned in the direction opposite to that of the ferromagnetic line defects on the Cu sites (except near the connecting 2-D structures).


The phase of MDWs immersed in an AFM background is suggestive of 
the stripes found experimentally by Tranquada et.~al.~\cite{Tranquada_stripes}.
However, unlike the stripe states,
these domain walls have an overall spin, and the AFM 
domains separated by the MDWs appear to be in phase.
Compared to theoretical studies,
the MDW state is reminiscent to the charged Bloch domain lines found in the seminal paper
of Zaanen \& Gunnarsson\cite{Zaanen_Gunnarsson} which studied the SSH model in 2--D, 
in that excess charge tends to localize on straight lines; however,
in the MDW state the domain lines are diagonal, ferromagnetic, and do not induce
a phase change across the domain line. Compared to the spiral state found by
Assaad \cite{Assaad-prb-93} at $U = 6$ and small doping, the MDW state is similar 
in that they both align along the diagonal (1,1)-direction. However, they differ in their properties.
The MDW, in some sense, may combine the presence of AFM spin modulations with ferromagnetism
as we will further discuss in \ref{ssec:intermediate}.

\begin{figure}[ptb]
\begin{center}
\includegraphics[width=7.0cm, angle=0]{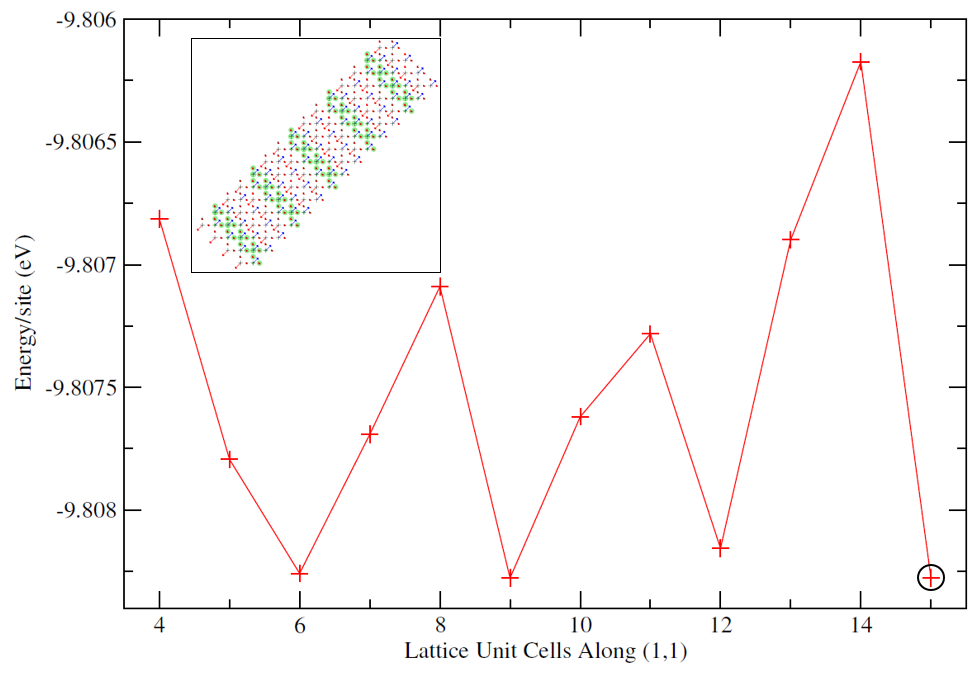}
\caption{
(Color online) Ground-state energy per site, $E/L$, vs.~supercell size 
along the diagonal $(1,1)$-direction for $4 \times L_{(1,1)}$ systems, for $ \Delta $ = 4.4, 
$h = 1/8$, tilted PBC. The inset shows the charge and spin order for the lowest $E/L$ solution 
(corresponding to the 
$4 \times 15$ supercell, indicated by a circle). Magnetic and charge orders are plotted in the same manner as in Fig.~\ref{fig:24x30}.
}
\label{fig:TiltEvL}
\end{center}
\end{figure}

Our calculations indicate that the MDWs 
tend to align periodically in one-dimension (1D) 
along the diagonal line, 
though there are
some two-dimensional features. We cannot rule out completely that 
these connecting structures 
arise solely from commensurability or boundary effects, 
since we have yet to find
a perfectly commensurate lattice that only contains 1D structures.
%
However, a clear preference for a periodicity in spacing is seen in the MDWs. 
To quantify this feature,
we scan long,
rectangular lattices. By varying the lattice in a single diagonal
direction, minima in the energy/site vs.~length will show which geometries are preferred for the
domain walls. Fig.~\ref{fig:TiltEvL} shows an example for $4\times L_{(1,1)}$ lattices.
A regular pattern is evident, 
with minima at  $L_{(1,1)}=6, 9, 12$, and $15$. 
Since there are two copper atoms per unit cell, this corresponds to 12, 18, 24, and 30 diagonal copper planes.
This suggests that, at $h=1/8$, the MDWs 
prefer lattices that allow regular spacing of 6 copper 
planes between the domain walls.

\begin{figure}[ptb]
\begin{center}
\includegraphics[width=7.0cm, angle=0]{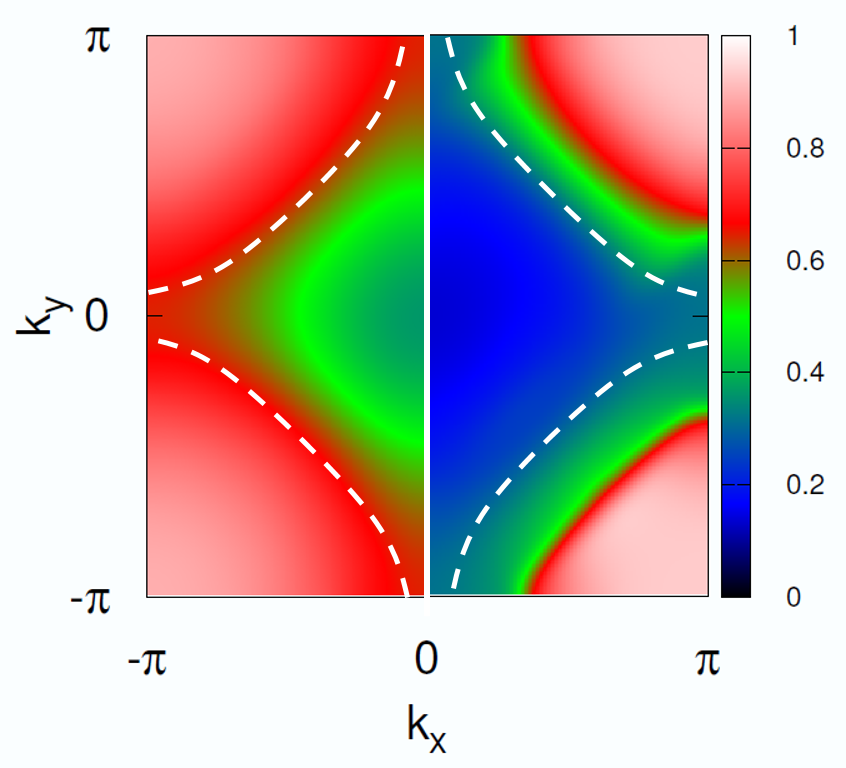}
\caption{
(Color online) Momentum distribution in MDW state. The total momentum distribution are shown 
for spin up (left) and spin down (right) particles from the $24\times 30$ supercell calculation 
of Fig.~\ref{fig:24x30}. The white dotted lines outline the Fermi surface of the non-interacting solution
for the same system.
}
\label{fig:phymomcombine}
\end{center}
\end{figure}

Fig.~\ref{fig:phymomcombine} depicts the total 
spin-$\uparrow$ and spin-$\downarrow$ 
momentum distributions $n_\sigma({\mathbf k})=
\langle \Psi| {\hat d}^\dagger_{{\mathbf k},\sigma} {\hat d}_{{\mathbf k},\sigma}+
\hat{p}^{\dagger}_{{\mathbf k},\sigma}\hat{p}_{{\mathbf k},\sigma} |\Psi\rangle$,
where ${\hat p}_{{\mathbf k},\sigma}$ is the  Fourier transform of  ${\hat p}_{j,\sigma}$ with 
$j$ running over both sets of basis vectors 
$\vec{r}_{{\rm O}_x}$ and
$\vec{r}_{{\rm O}_y}$, and similarly ${\hat d}_{{\mathbf k},\sigma}$ is the Fourier transform of 
${\hat d}_{i,\sigma}$.
We see the spin imbalance
present in the system, as well as the tendency to develop diagonal modulations. 
Symmetries about the $k_x$ or $k_y$ axes are both broken. 
In the figure, the left half shows $n_\uparrow({\mathbf k})$ while the right half shows 
$n_\downarrow({\mathbf k})$. The missing portions of the momentum distributions can be constructed 
by reflection with respect to the origin, $n_\sigma(-{\mathbf k})=n_\sigma({\mathbf k})$
In addition to the broken spin symmetry, we see significant re-construction of 
the Fermi surface to create nesting which produces the modulated MDW structures.
We comment in passing 
that spin imbalance with attractive interactions, for example
in Fermi atomic gases on an optical lattice, 
are expected to have non-trivial modulated pairing states. It would be interesting to investigate 
possible relation between the spin imbalance seen here and potential non-trivial pair density wave states, 
using more advanced many-body methods in future studies.

\subsubsection{Formation of Domain Walls from Low Doping}
\label{ssec:formDW}

\begin{figure*}[ptb]
\centering
  \subfloat[h=1/72]{%
    \includegraphics[width=5.5cm]{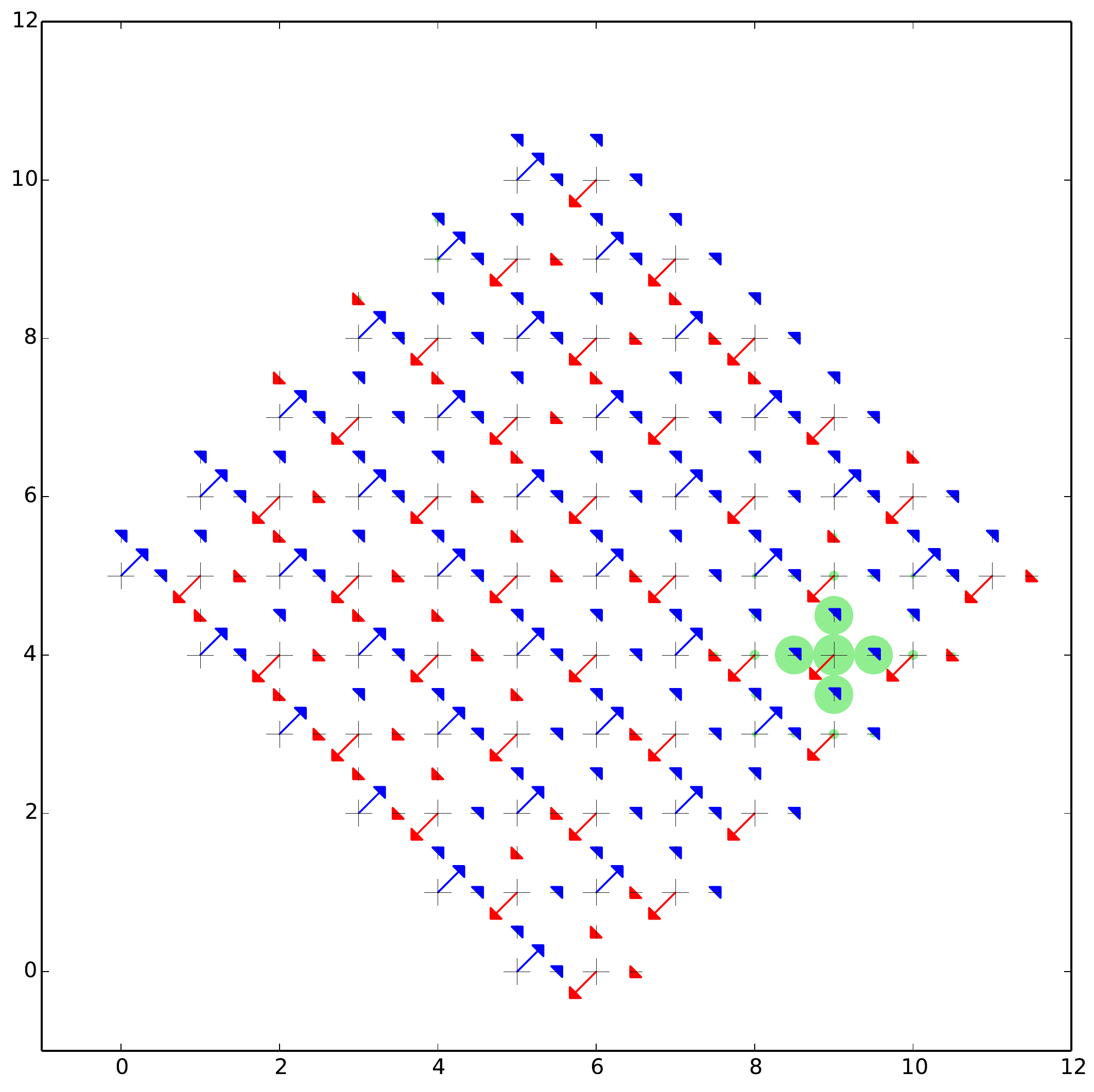}}\hfill
  \subfloat[h=1/36]{%
    \includegraphics[width=5.5cm]{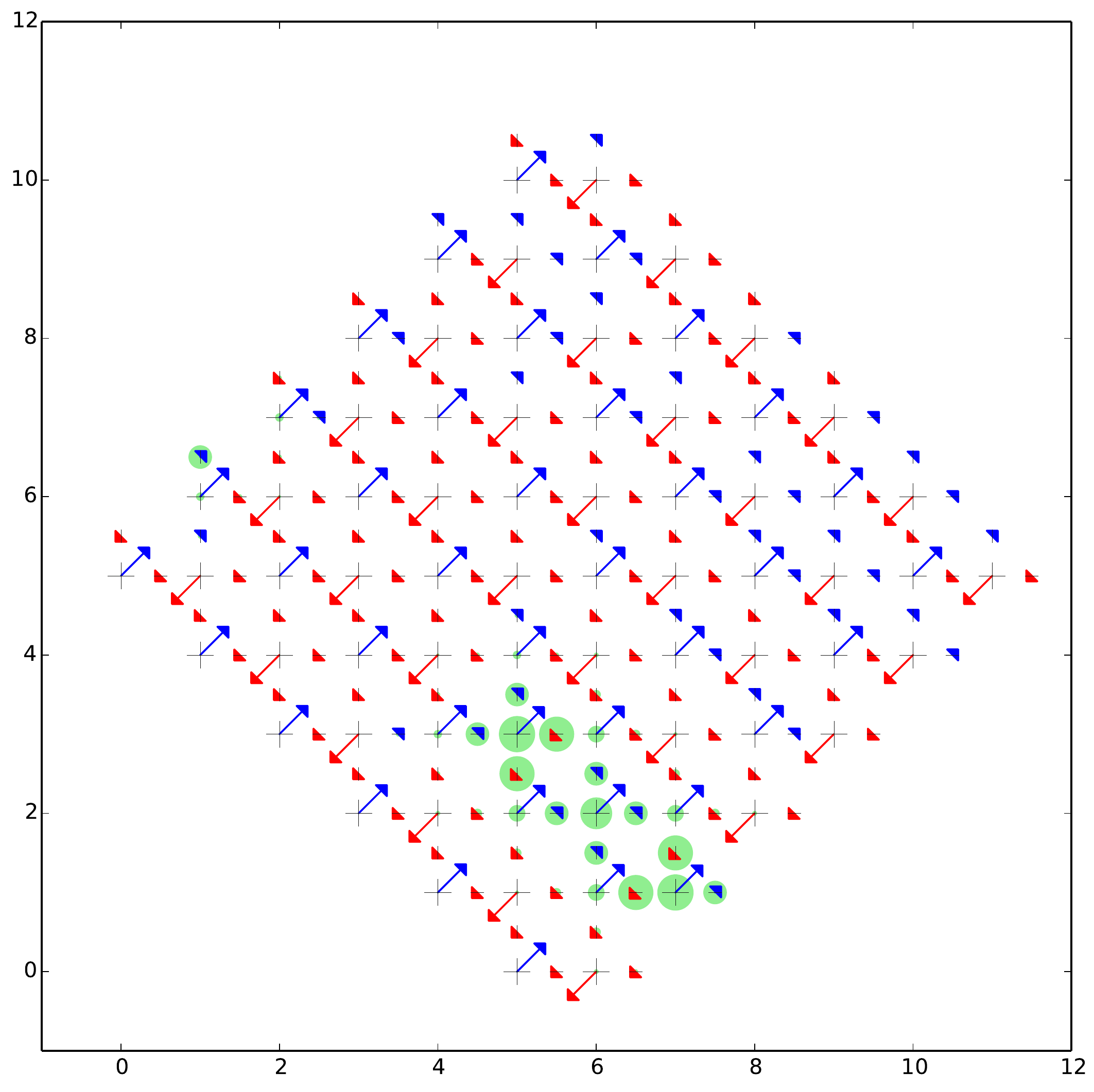}}\hfill
  \subfloat[h=1/36]{%
    \includegraphics[width=5.5cm]{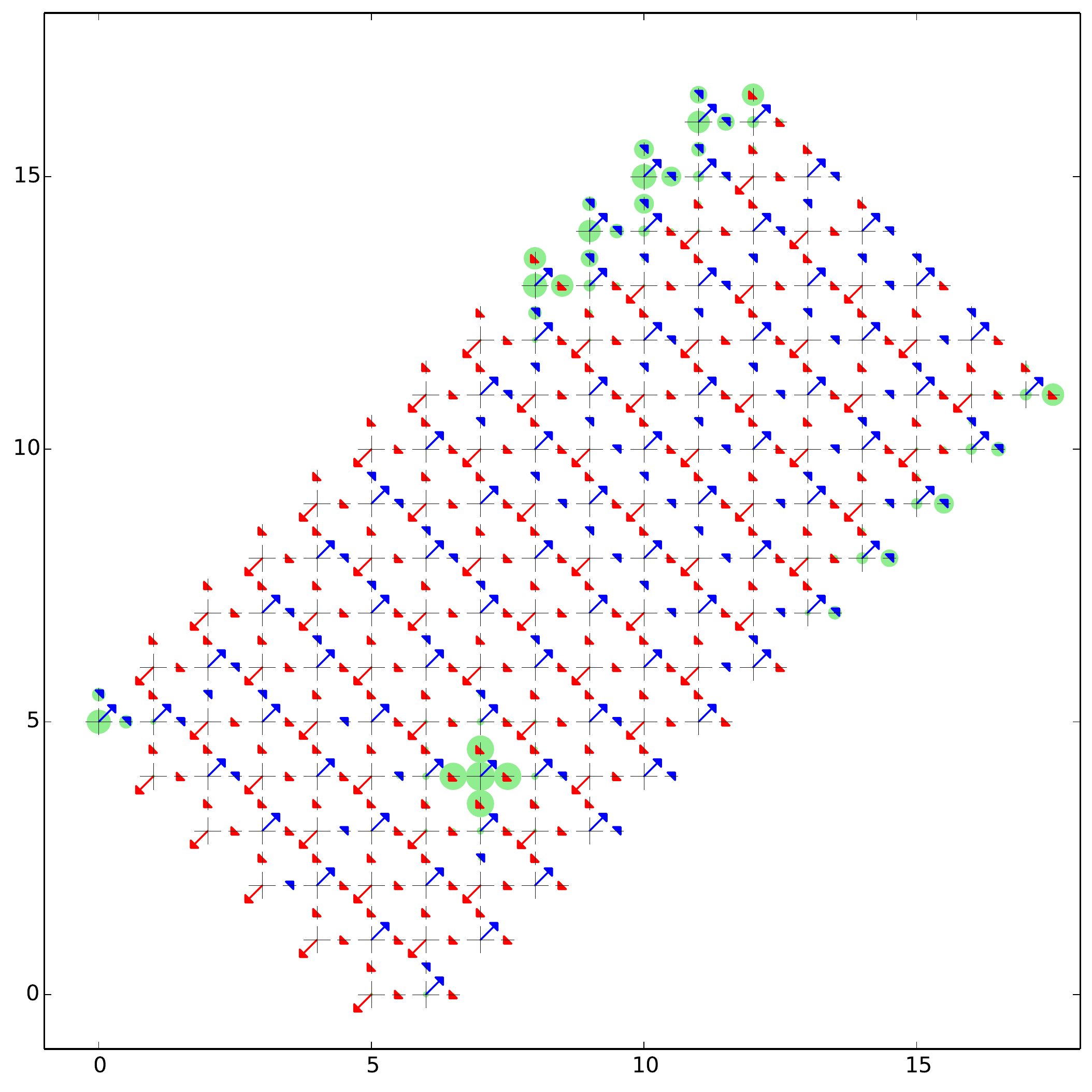}}
  \caption{
    (Color online) Formation of MDWs from low doping, at $\Delta=4.4$. 
    The crosses represent the $d$-orbital on the copper atom
     and the horizontal and vertical dashes represent
    the $p_x$ and $p_y$ orbitals on the oxygen, respectively. The colored arrows are proportional to the spin
    (which are all oriented in one arbitrary direction as in Fig.~\ref{fig:24x30}),
    with blue representing positive spin and red representing negative.
    The green circles  indicate  where the extra charge (holes) localize, with their sizes
    proportional to the excess hole density on a given site ($\delta n_{\rm excess}$).
    Tilted supercells with PBC are used in the calculations, with sizes: 
    a) $6\times 6$ (single hole doped); 
    b) $6\times 6$ (two hole doped); 
    c) $6\times 12$ (four hole doped). 
  }
  \label{fig:hole_scan}
\end{figure*}

To probe the mechanism that leads to 
the formation of the 
MDW within the AFM background, we scan lightly doped systems.
The plots in Fig.~\ref{fig:hole_scan} illustrate how a small number of holes
accumulate and nucleate. 
For single hole systems, the calculations show that the extra hole (green circle) localizes around a randomly
chosen copper site in a periodic supercell. 
There is some smearing over to local O sites as well as nearest neighbor
copper sites. 
The system has perfect AFM order on the copper atoms, except for the
site where the hole localizes. On that site, there is local ferromagnetic order with neighboring copper
sites. The direction of the spin on the neighboring $p$-bands oppose this ferromagnetic order.
This behavior is very robust, and is seen in all our calculations with a single hole under PBC, 
for all lattice sizes and geometries.
We call this local magnetic and charge order a spin flip defect.

We interpret the spin flip defect, in which the extra hole density is accompanied by an overall spin flip 
on the corresponding
Cu site,
 as a way to gain exchange energy. 
The density increase within the ferromagnetic defects (see 
Sec.~\ref{ssec:magDW}) makes the total density of the 
 CuO$_2$ unit cell well above unity (indeed higher than $1+h$) along the MDW. This renders the 
 AFM state, which avoids double occupancy of the $d$-sites, less effective. The system then chooses
to flip the spin on the $d$-site and introduce local ferromagnetic order on the Cu sublattice. 
In order for this exchange energy gain to be efficient, the excess hole has to remain localized around the spin-flipped site, consistent with the numerical results.
 If these defects can be properly connected along the diagonal 
line with minimal frustration,  the gain in exchange energy is maximized.

When the system is doped with two holes, the ground state under PBC 
shows three spin 
flip defects that bind into a diagonal line as can be seen from the local ferromagnetic behavior
superimposed on the AFM 
background.
This diagonal line of three spin flips contains all of the extra hole density.
\COMMENTED{
Different from the 1-hole doped system above, the spin on the Oxygen $p$-sites tended to weakly align with, rather than oppose, the ferromagnetic order on the Copper $d$-sites.
Exceptionally, we can see on the $p$-sites between the spin flipped  $d$-sites opposed the ferromagnetic order, 
and the excess charge on those $p$-sites is significantly greater than that on the sites that aligned.
\REMARKS{What is the significance of this? What is energetically gained from this?}
}
We will refer to this as a magnetic line defect from  the AFM order.
In less energetically favorable systems, two spin flips similar to the ones in the
one hole systems were seen. 
Our calculations on these systems 
suggested that the single 
spin 
flip defects repel each other, 
and the lowest energy state is reached when they bind into a magnetic line defect occuring over three diagonal Cu sites. 

The last system in Fig.~\ref{fig:hole_scan} has the same doping, $h = 1/36$, as the two-hole doped system
just discussed;
however, 
it has a supercell twice the size, thus giving a doping of four holes.
The calculations show the magnetic line defects growing longer 
(5 Cu sites vs.~3) with similar spin and charge order on the O $p$-sites as the two-hole doped system. 
Along with the magnetic line defect is an isolated spin flip defect.
The calculations suggest that the system slowly builds 1-D magnetic  line defects as doping is increased.
At first the excess holes repel until 
a sufficient number of defects are present and it becomes beneficial to combine them.
(Pan and Gong\cite{Pan_3band} had performed mean-field studies of a few-hole doped systems, 
though they did not explore a similar parameter space.)

These calculation suggest the following physical picture from GHF. At very low doping 
there are isolated holes and short magnetic defect lines. As doping increases,
the lines start to create one-dimensional MDWs 
which are spaced away from each other.
Around $h=1/8$ doping, 
approaching the maximum superconducting transition temperature,
closely spaced, mostly one-dimensional, domain walls dominate the system. As doping is further 
increased,
the system starts to create orthogonal domain walls, eventually creating checkerboard patterns.
Some of these features are likely related to the situation in the intermediate $\Delta$ regime,
which is discussed below in Sec.~\ref{ssec:intermediate}.

\COMMENTED{
It will be very 
interesting, in future studies, to investigate possible connections of these characteristics
to superconducting order.
}

\subsection{Spin Spirals}
\label{ssec:spirals}

\subsubsection{Magnetic and Charge Order}

We now turn to the lowest considered charge-transfer energy, $\Delta = 1.5$ $eV$.
In this regime, we find an AFM state at half-filling with $n_d \approx 0.4657$ and
$n_p \approx 0.2672$.
At 
doping of $h=1/8$, we find a 
new spin order, 
characterized by a uniform charge distribution with a long wavelength planar spin spiral.
In the GHF ground state, 
most of the extra charge (holes) are localized on the oxygen $p$-orbitals, 
as opposed to the 
$d$-orbitals on the Cu sites.

\begin{figure*}[ptb]
\centering
  \subfloat[]{%
    \includegraphics[width=6.0cm, angle=0]{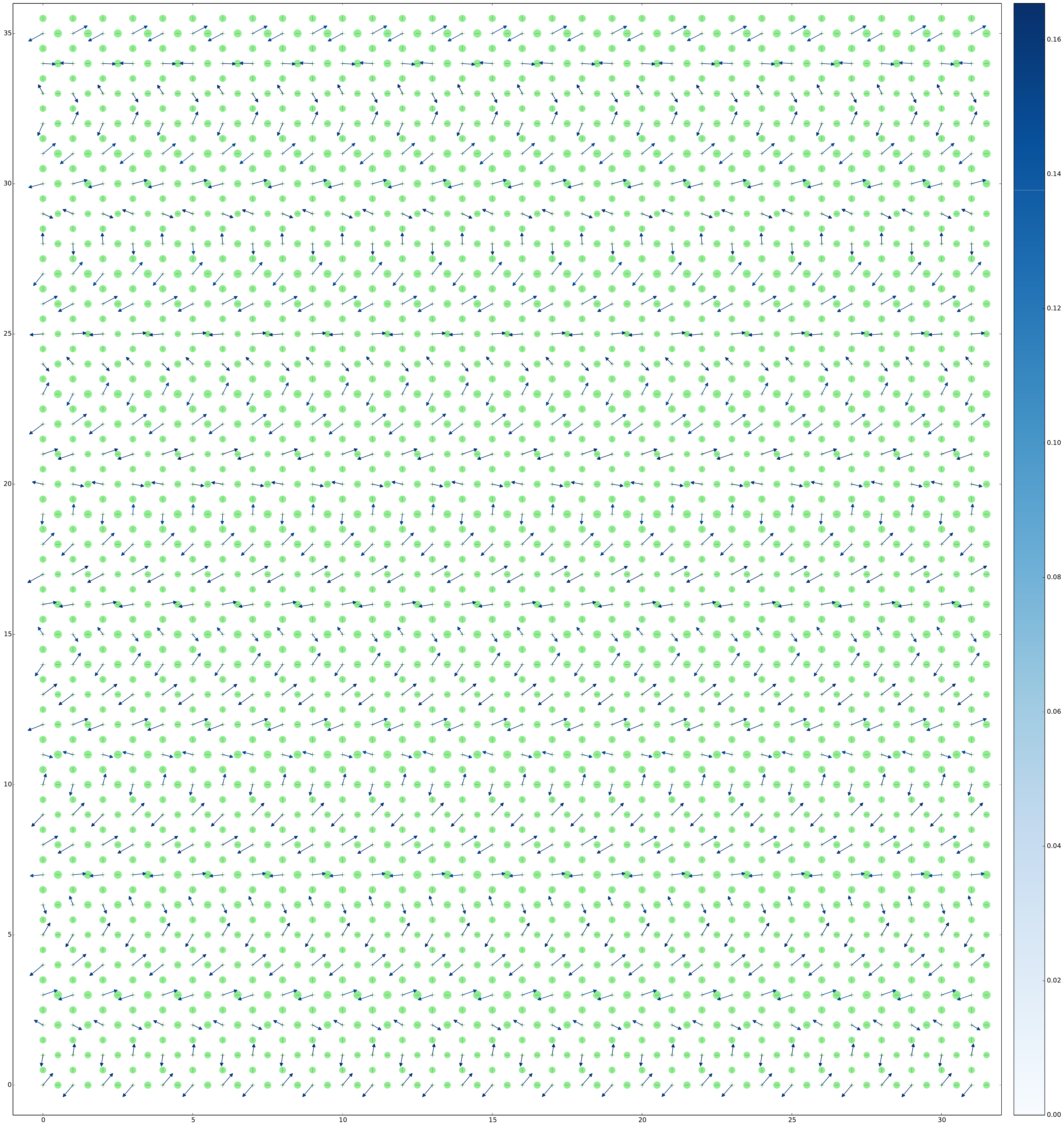}}\ \ \qquad
  \subfloat[]{%
    \includegraphics[width=3.0cm, angle=0]{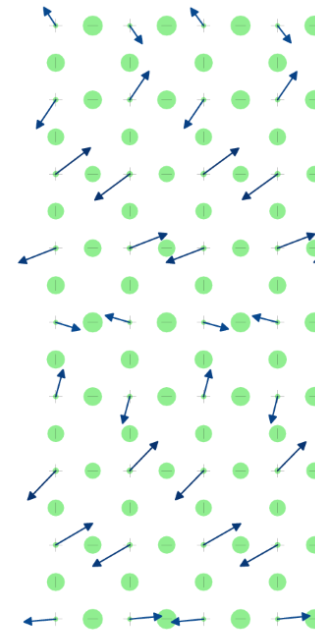}}\ \ \qquad
  \subfloat[]{%
    \includegraphics[width=6.0cm, angle=0]{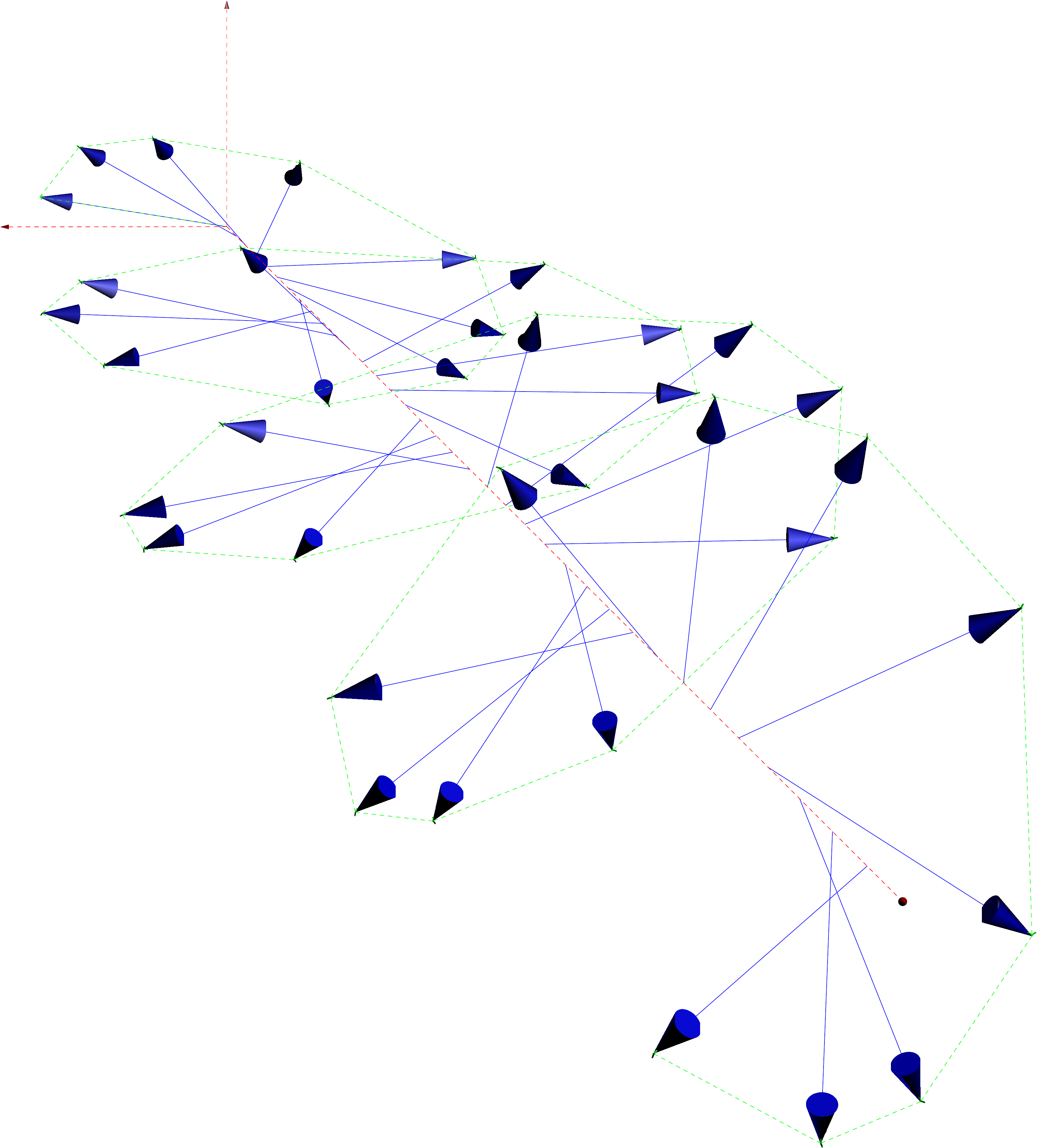}}
  \caption{
  (Color online) Charge and magnetic order for a periodic $32 \times 36$ supercell,
with $ \Delta $ = 1.5 eV and $h = 1/8$.
In (a), the spins (arrows) are plotted as a projection in the $x$-$y$ plane. 
The total spin, $S_{tot}(\vec{r})$, on each site is given by the color gradient on the right. 
The spin on the O sites is negligible.
The size of the green circles are proportional to the excess hole density from half-filling, $\delta n_{\rm excess}(\vec{r})$.
A small section of the lattice can be seen in (b), better highlighting the order.
In (c) a 3-D plot of the staggered spin is shown along a line cut at $x = 0$, viewed along the $y$-axis. Blue arrows are the real 3D staggered spin and green dashed lines connect
near neighbor spins to highlight spiral structure.
}
\label{fig:32x36}
\end{figure*}

In Fig.~\ref{fig:32x36}, we show the spin density 
($S_x(\vec{r}),S_y(\vec{r}), S_z(\vec{r})$) 
and the excess hole density, $\delta n_{\rm excess}(\vec{r})$  on a lattice containing $32 \times 36$ unit cells with periodic
boundary conditions, that is, from a ${\rm Cu}_{1152} {\rm O}_{2304}$ supercell. 
The spin on the $p$-orbitals is negligible and is omitted in the plot.
The charge order is uniform across the lattice.
The leading spin order is anti-ferromagnetic on the Cu sites; however, the spin is slowly turning in a randomly chosen
plane.
 Unlike in a linear spin-density wave, the spiral has a near constant total spin, $S_{tot}$. 

In the right panel of Fig.~\ref{fig:32x36}, an image of the 3--D \emph{staggered} spin density, $(-1)^{x_i+y_i}\,{\vec S_i}$ is shown along a line cut. 
The spin rotation is almost perfectly constrained to a
 plane 
(i.e., if all spins are translated to a single point, the spin vectors lie in a plane), with 
the orientation of the plane seemingly random.
The projection of a spin spiral  onto a single spin orientation, which would be 
typically how the spins are resolved experimentally, would appear as a linear spin wave or AFM domains.
Our results suggest that such structures could be an indication of a more complex
three-dimensional spiral behavior.

\COMMENTED{
The top portion of Fig.~\ref{fig:4x36} is a spin and charge order plot.
The colored arrows are proportional to spin
in the dominant direction (Sz in this case), with positive values colored red and negative values
colored blue. The green circles are proportional to the difference in density as compared
to the same system with anti-ferromagnetic order at half-filling.
Fig.~\ref{fig:4x36} would suggest an anti-ferromagnetic spin wave in Sz along y
with a wavelength of 9 copper sites, perfect AFM order along x, and a charge wave along y characterized
by extra holes concentrated on the p-bands with a wavelength of 4 copper sites.
The bottom of the figure is a 3-D representation of the overall staggered spin which highlights
the AFM spiral along the line cut highlighted by the red box. The blue arrows are the staggered
spin on the copper site. The green line connects the tips of neighboring arrows.
Fig.~\ref{fig:long-stag} is the same staggered plot viewed along the y-axis to better highlight
the  planar spiral.}

\begin{figure}[ptb]
\begin{center}
\includegraphics[width=7.0cm, angle=270]{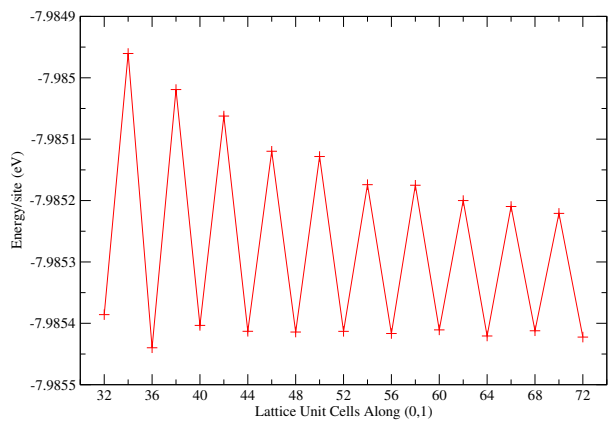}
\caption{
(Color online) Energy per site vs. lattice length along y for $4 \times L_y$ systems at $ \Delta $ = 1.5 eV, $h = 1/8$, PBC.
}
\label{fig:EvL_nt}
\end{center}
\end{figure}

\COMMENTED{
The order 
is very robust against changes in geometries and sizes. The spiral
always propagate 
along the copper-oxygen bonds.
The spiral order has a wave-vector along $y$-direction, breaking $x$ and $y$ symmetry, with a wavelength of approximately 9 Cu sites.
}

To estimate the wavelength of the spiral, we investigated the dependence of the energy
per site, $E/L$, on $L_y$ in $4 \times L_y$ supercells.
As we show in Fig.~\ref{fig:EvL_nt},
there are comparable energy solutions with spiral 
wavelengths of 7, 8, 9, and 10 Cu sites. The best solutions, as indicated by 
$E/L$, occur at $L_y=36,72$, which would suggest a spiral wavelength of 9 Cu sites.
On the other hand, perfect spiral order with constant charge order were also observed
in an $8 \times 8$ supercell . 
Further,
for wider $8 \times L_y$ supercells, the spiral tends to align along the (short) $x$-direction. 
We then performed a comprehensive study of $16 \times 18$ systems, which shows a variational energy preference to align the spiral in the $y$-direction ($L_y=18$) rather than $x$. 
This leads us to believe the spiral wavelength is closer to 9 Cu sites, with uniform charge density.
The order 
is very robust against changes in geometries and sizes. The spiral
s always oriented along $x$- or $y$-direction, propagating 
along the Cu-O 
bonds.

\subsubsection{Momentum Distributions and Nesting}

\begin{figure*}[ptb]
\centering
  \subfloat[]{%
    \includegraphics[width=8.0cm, angle=0]{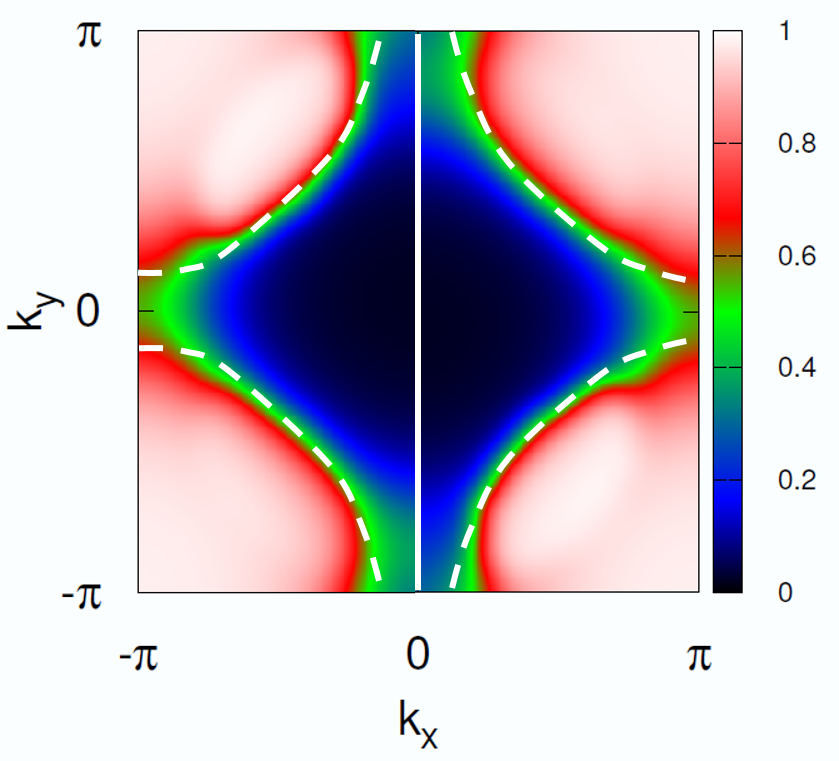}}\ \ \qquad \qquad
  \subfloat[]{%
    \raisebox{1.0cm}[0pt][0pt]{
    \includegraphics[width=6cm, angle=0]{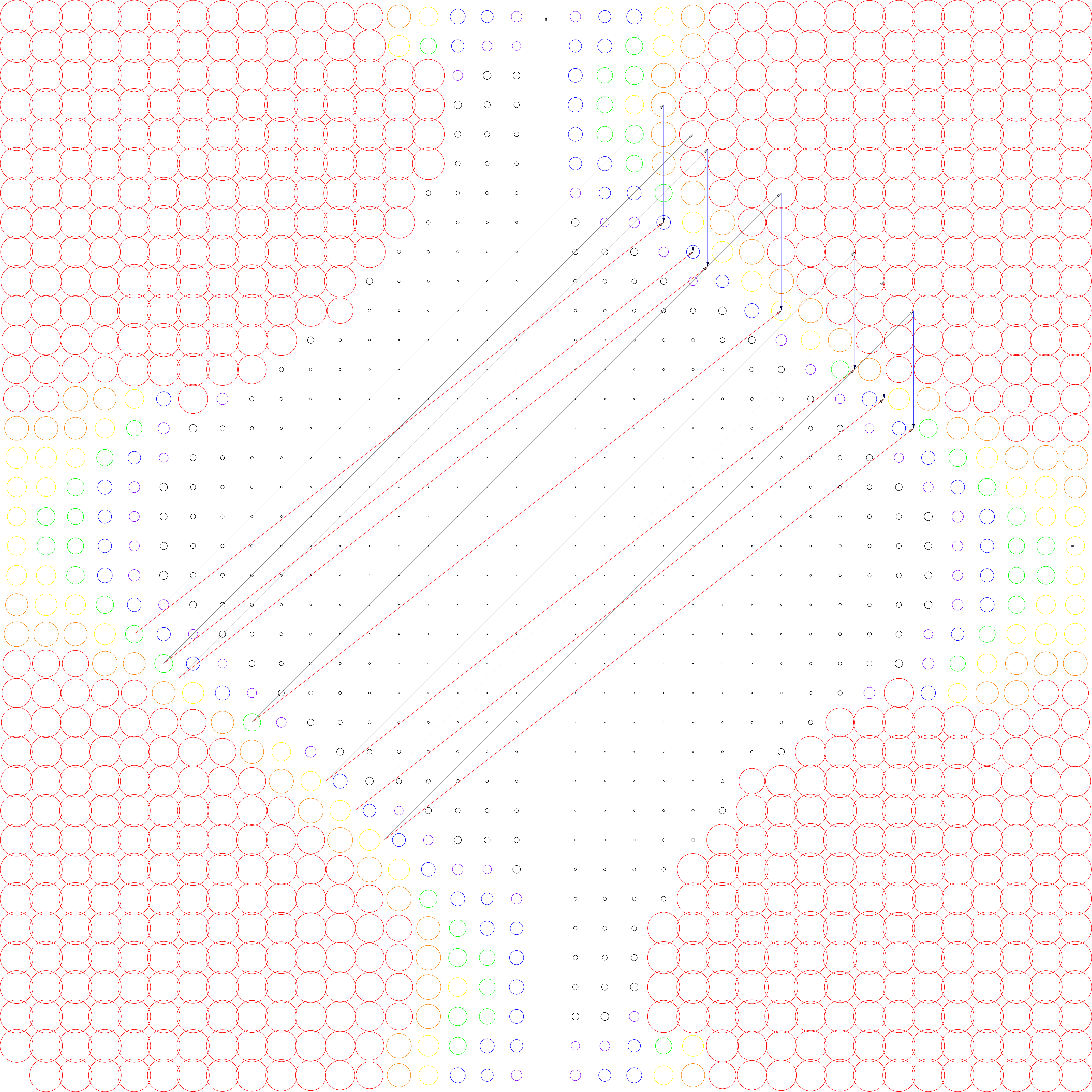}}}
  \caption{
  (Color online) Momentum distribution and nesting structure for spin spiral states. 
  The left panel shows $n_\sigma({\mathbf k})$ for $\sigma=\uparrow$ (left half) and $\downarrow$ (right half)
  in the $32\times 36$ supercell in Fig.~\ref{fig:32x36}.
  The color scale is set to show
nearly fully occupied momenta in white/red, nearly empty momenta in blue/black, and half occupied
momenta around the Fermi surface in green.
The white dotted lines outline the Fermi surface of the non-interacting solution for the same system.
  The right panel illustrates the nesting structure.
 Similar to the left panel, $n_\uparrow$ is on the left half and $n_\downarrow$ is on the right.
The size of the circles is proportional to the occupation (magnitude of $n_\sigma$). The colors are to guide the eye around the
Fermi surface. 
  The black lines represent a ($\pi, \pi$) vector. 
  The red lines connect complementary points ${\mathbf k}$ and ${\mathbf k'}$
  between $n_\uparrow({\mathbf k})$ and $n_\downarrow({\mathbf k'})$ 
  near the Fermi surfaces (see text).
The blue lines show the shift  $({\mathbf k'}-{\mathbf k})$
[with respect to ($\pi, \pi$)], which defines the spiral wave vector.
}
\label{fig:32x36_mom}
\end{figure*}

\COMMENTED{
\begin{figure}[ptb]
\begin{center}
\includegraphics[width=7.0cm, angle=0]{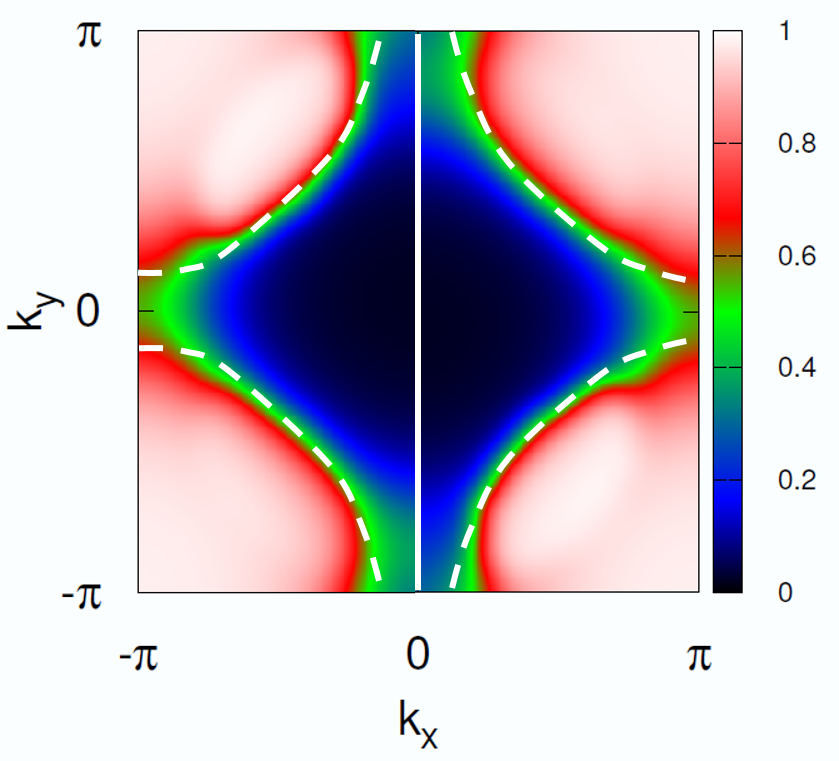}
\caption{
(Color online) Momentum distribution for spin up (left) and spin down (right) holes in the 32x36 lattice.
}
\label{fig:32x36_mom}
\end{center}
\end{figure}

\begin{figure}[ptb]
\begin{center}
\includegraphics[width=7.0cm, angle=0]{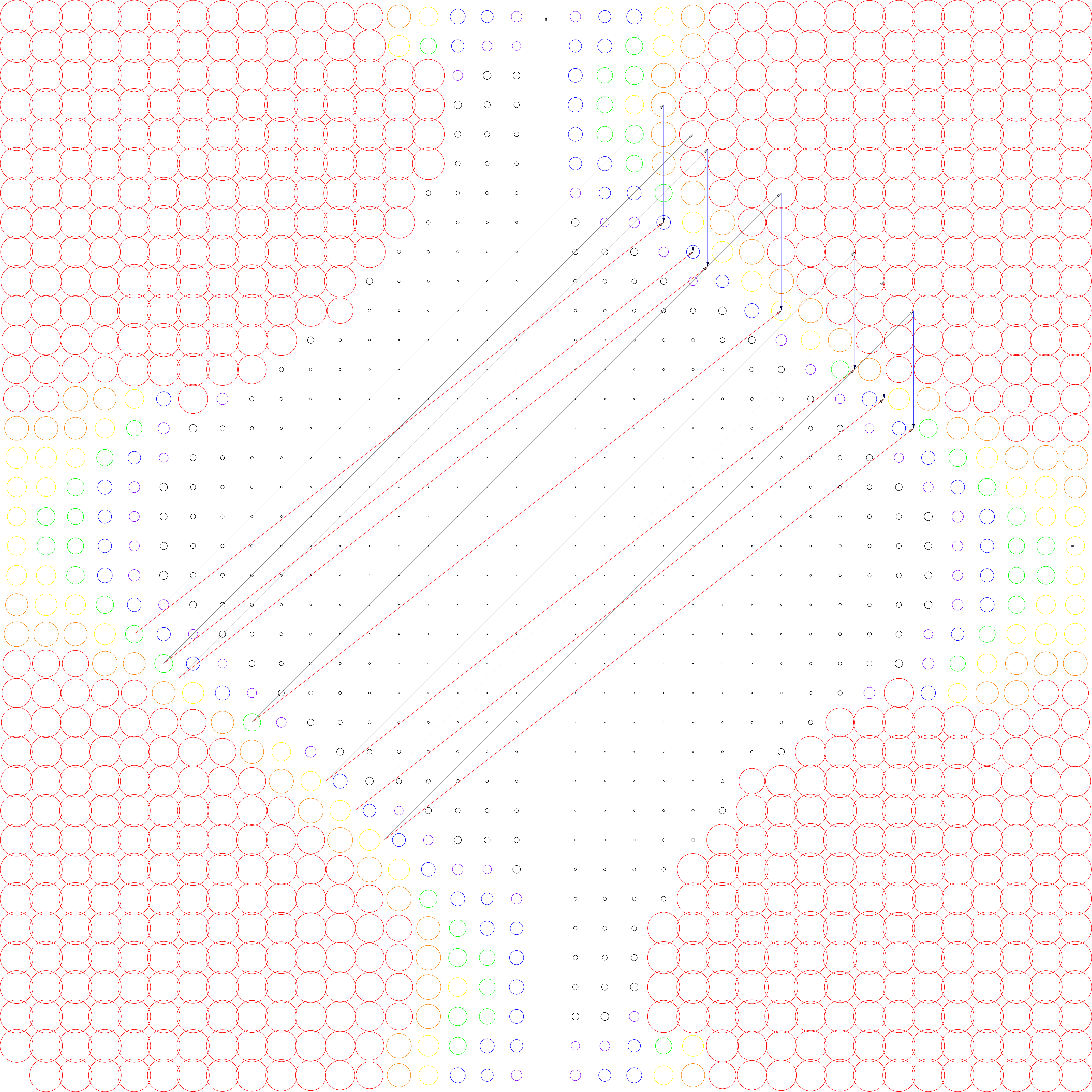}
\caption{
(Color online) Combined momentum distribution plot.
Spin up distribution is on the left and the down on the right.
The black lines represent a ($\pi, \pi$) vector. The red lines connect complementary points on the Fermi surfaces.
The blue lines show the shift from ($\pi, \pi$).
}
\label{fig:36x36_combine}
\end{center}
\end{figure}
}

In this section, we examine the properties of the GHF ground state in momentum space. 
In particular, we are interested in detecting and understanding 
nesting properties of these systems.
As in Sec.~\ref{subbssec:magDW-1o8}, we consider spin-resolved momentum distributions 
$n_{\uparrow}(\mathbf{k})$ and  $n_{\downarrow}(\mathbf{k})$.
Fig.~\ref{fig:32x36_mom} shows 
the momentum distributions 
for a large, $32 \times 36$ supercell. Only half of $n_\sigma(\mathbf{k})$ is displayed 
for each $\sigma$.
In the GHF solution, symmetry is broken along $y$ but preserved in the $x$ direction,
so that the full momentum distribution can be recovered by 
reflection with respect to the $k_y$ axis, $n_\sigma(-k_x,k_y)=n_\sigma(k_x,k_y)$.
The re-construction at the Fermi surface is evident relative to the non-interacting momentum distribution.
(It should be stressed that all the formulation is in terms of holes; 
the momentum distributions for electrons can of course be mapped straightforwardly from these results.)

\COMMENTED{
There is an obvious symmetry
breaking within and between the two distributions. In both distributions, there is symettry about the $k_y$ axis;
however, there is no symmetry along the $k_x$ axis or either diagonal axes. The Fermi surface is distrorted in both distributions from the half-filled case.
Also notice that the distributions are mirrored images of eachother about the $k_x$ axis.}

We observe a spectacular 
interplay between $SU(2)$ symmetry breaking and translational
symmetry breaking, resulting in a nesting property that stabilizes a spin spiral.
For perfect AFM order, the nesting vector, $\vec{Q}$, should coincide with $(\pi , \pi )$. A shift in this nesting vector corresponds to an instability towards a modulated phase.
In the one-band Hubbard model, for example, the HF solution is found \cite{Xu-JPCM-11}
to produce linear spin density 
waves in large portions of the parameter space, 
the symmetry $n_\uparrow({\mathbf k})=n_\downarrow({\mathbf k})$ is preserved, and the 
spin density waves can be thought of as a linear combination of two counter-propagating 
spirals.
Here we have a broken symmetry in the momentum distribution between $\uparrow$- and $\downarrow$-spins, 
resulting in a non-colinear spin wave.
The right panel of Fig.~\ref{fig:32x36_mom}
illustrates the nesting more quantitatively.
We identify complementary points ${\mathbf k}$ and ${\mathbf k'}$ near the two Fermi surfaces for which
 $n_\uparrow({\mathbf k}) + n_\downarrow({\mathbf k'}) \doteq 1$ within a few percent.
The large number of pairs found indicate that the spiral is created predominantly by a simple pairing mechanism \cite{Zhang-Ceperley-PRL-08,Xu-JPCM-11} involving two primary planewaves.
[The simplest model \cite{Overhauser} to help visualize the spiral state is spin orbitals of the form 
$u\,\exp(i\,{\mathbf k}\cdot {\mathbf r})|\uparrow\rangle + v\,\exp(i\,{\mathbf k'}\cdot {\mathbf r}) |\downarrow\rangle$, 
where $|u|^2+|v|^2=1$ and the nesting vector is ${\mathbf Q}\equiv {\mathbf k'}-{\mathbf k}$.]
From the figure we see that the nesting vector is
consistently shifted along the $y$-direction
by $4 \times (2\pi)/36$, which corresponds to 9 Cu sites in real space.
This is consistent with the numerical estimation above.

\subsubsection{Behavior at Low Doping}

We also explored the cases with very low doping (one and two doped holes) in a periodic lattice,
as we have done for the magnetic domain wall phase in Sec.~\ref{ssec:formDW}.
For an $8 \times 8$ supercell, Cu$_{64}$O$_{128}$, at $h=1/64$ and $1/32$ doping, the order of the systems showed very little deviation 
from half-filling. For a single  hole ($h=1/64$),
the calculations show that the order is nearly a perfect anti-ferromagnet. There are some minor periodic 
modulations of the total spin in the Cu $d$-orbitals from an AFM background.
The density is also very close to uniform, with minor modulations coinciding with the Cu
spin deviation. 
As in the case of $h=1/8$, most of the extra density
lies on the O $p$-orbitals. For two holes ($h=1/32$), the calculations show similar order as the single  hole case. 
The spin shows near perfect AFM order 
with periodic deviations 
in the total spin. Compared to the single hole system, the periodic deviation 
occurs twice as often. The charge order is nearly uniform with most of the extra hole density
on the O $p$-orbitals, again with minor modulation coinciding with that of the total spin.

\begin{figure*}[ptb]
\centering
  \subfloat[2.0eV]{%
    \includegraphics[width=5.5cm]{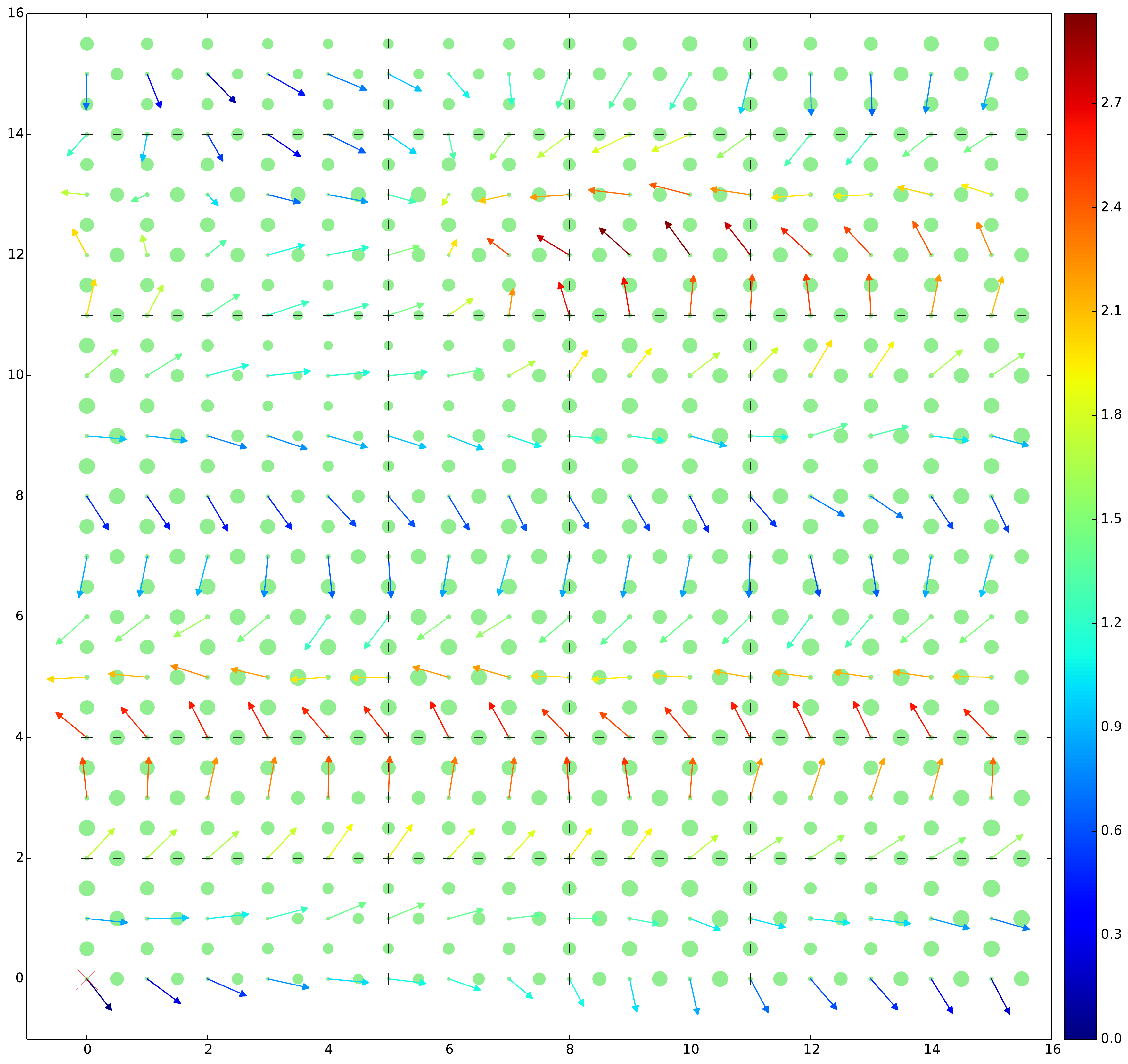}}\hfill
  \subfloat[2.1eV]{%
    \includegraphics[width=5.5cm]{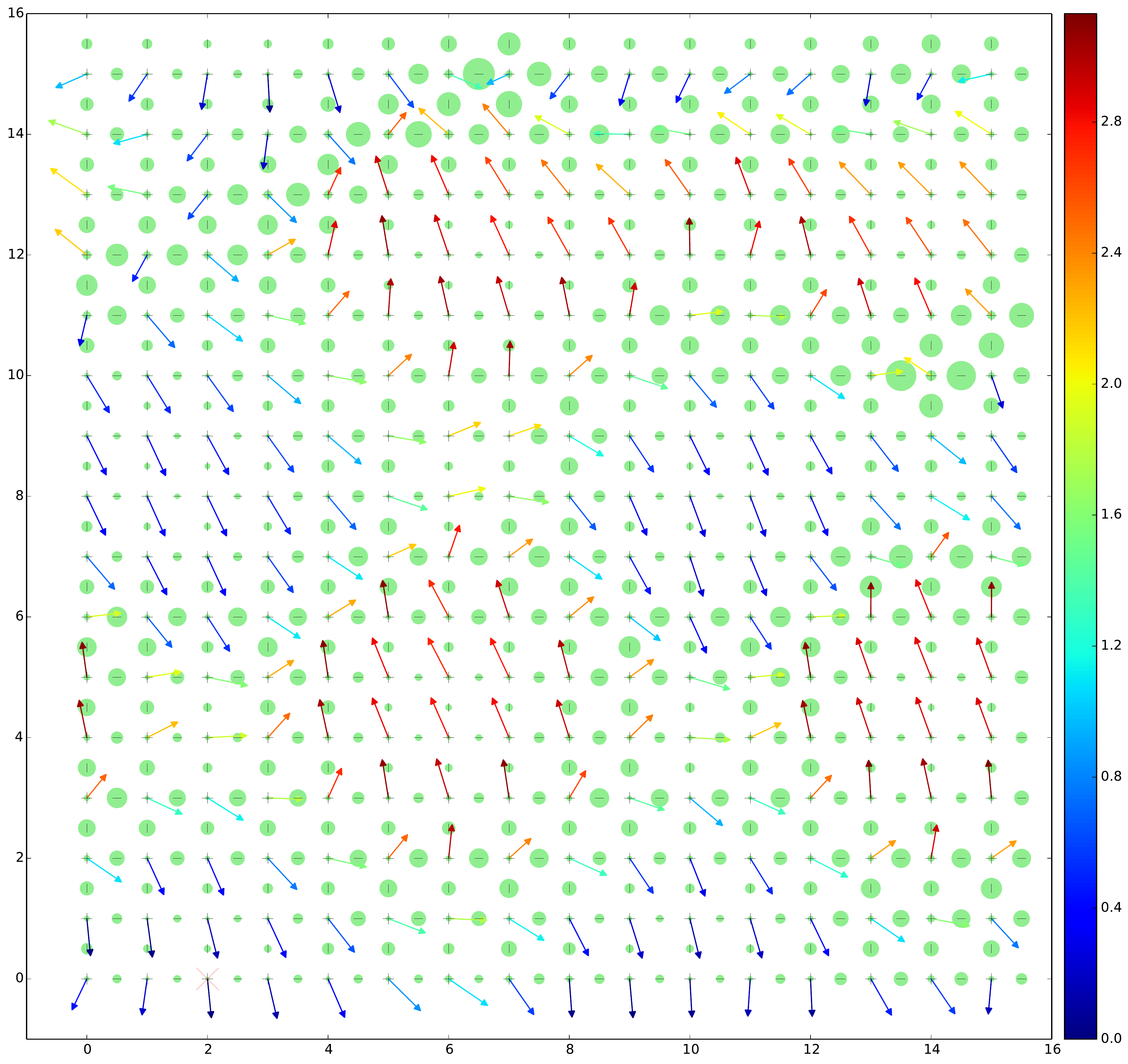}}\hfill
  \subfloat[2.2eV]{%
    \includegraphics[width=5.5cm]{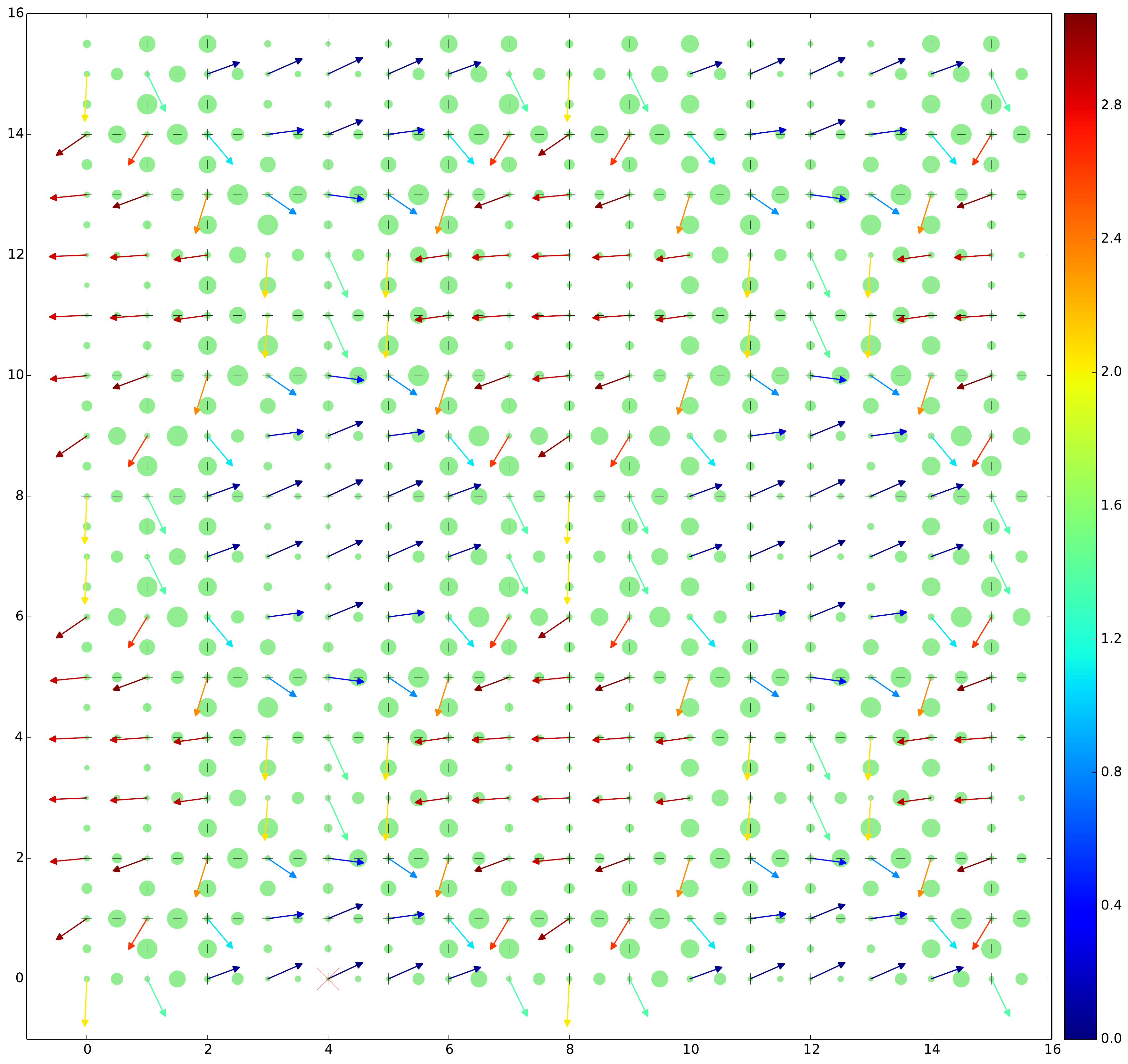}}\\
  \subfloat[2.4eV]{%
    \includegraphics[width=5.5cm]{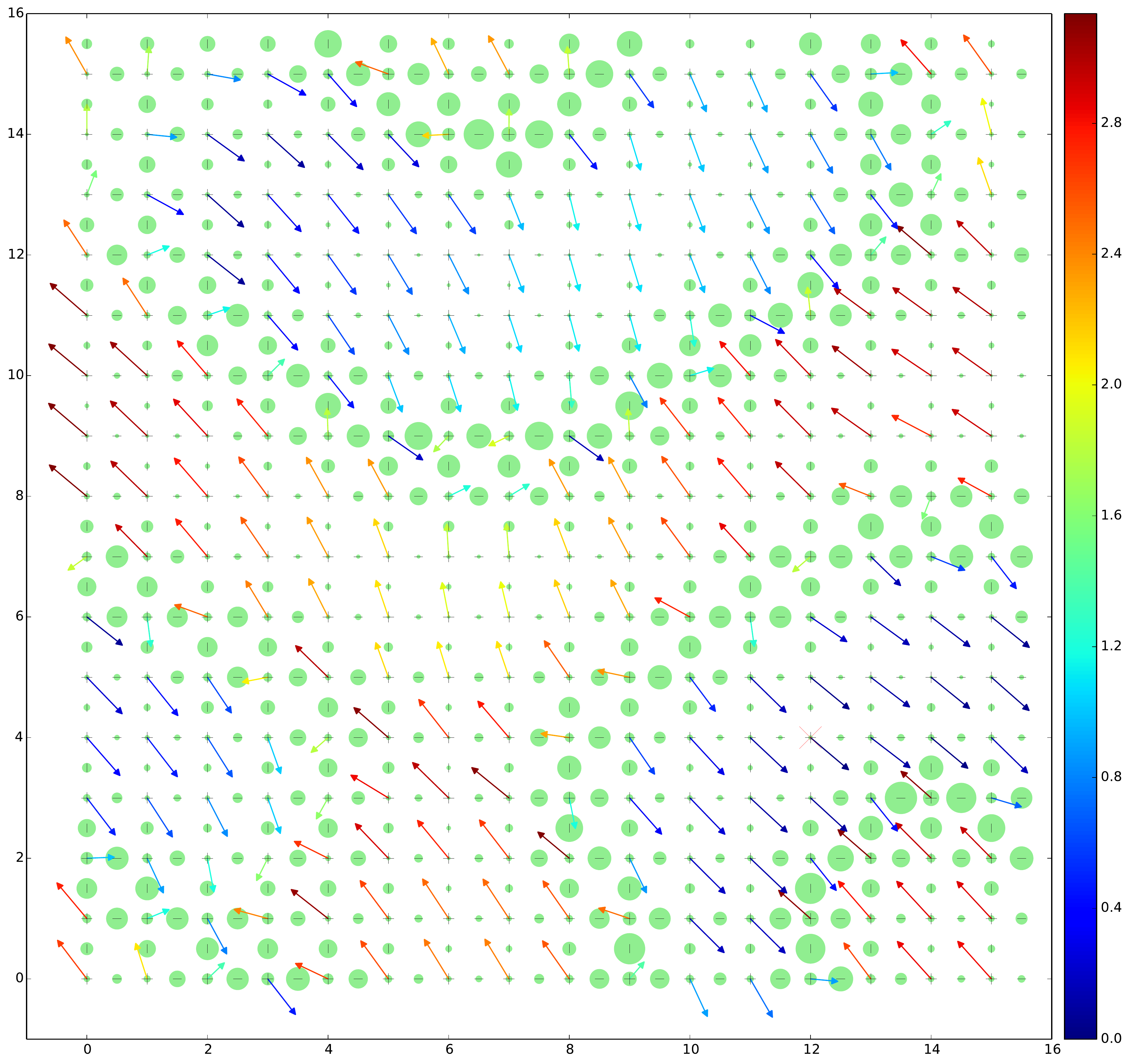}}\hfill
  \subfloat[2.5eV]{%
    \includegraphics[width=5.5cm]{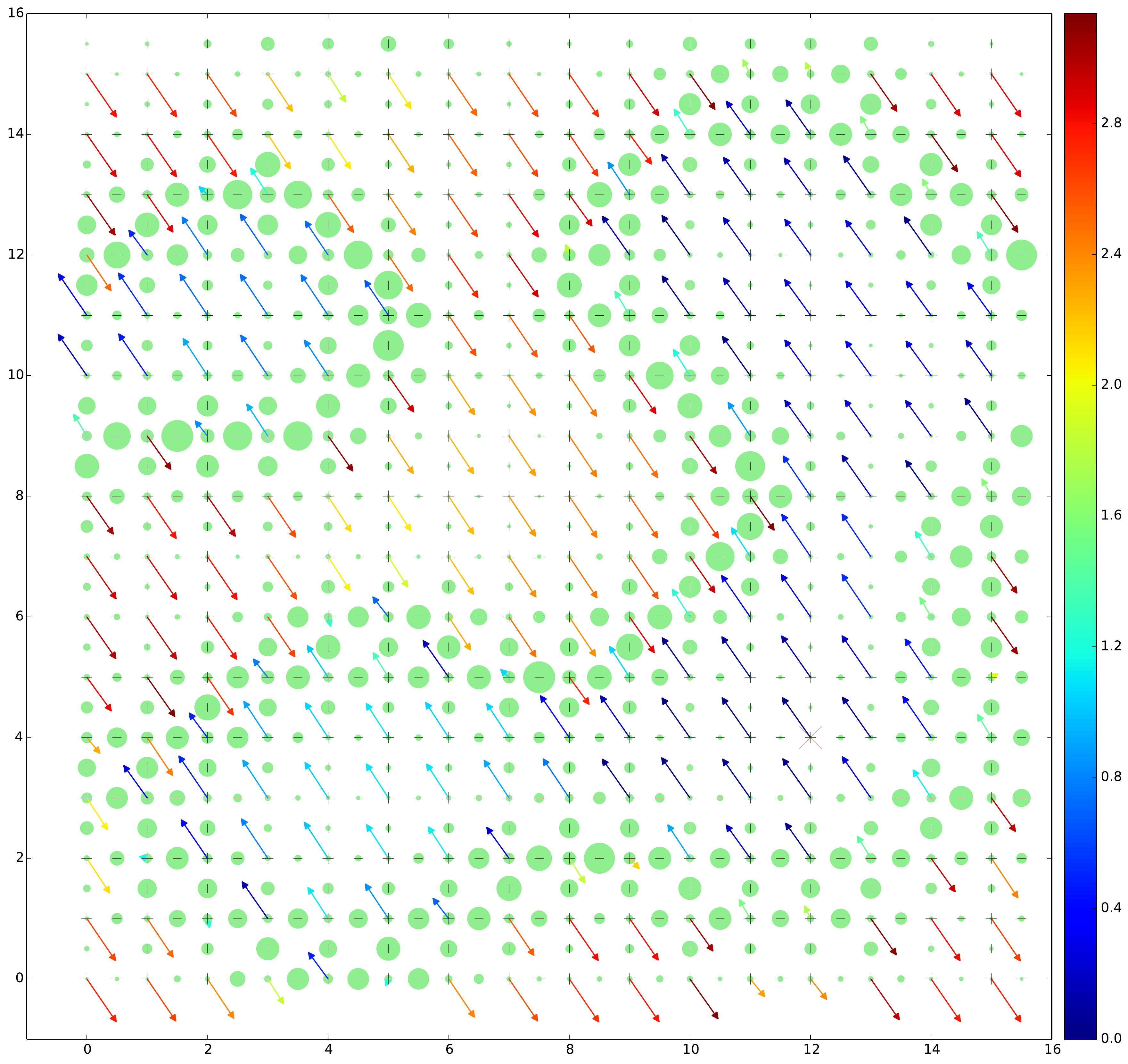}}\hfill
  \subfloat[2.6eV]{%
    \includegraphics[width=5.5cm]{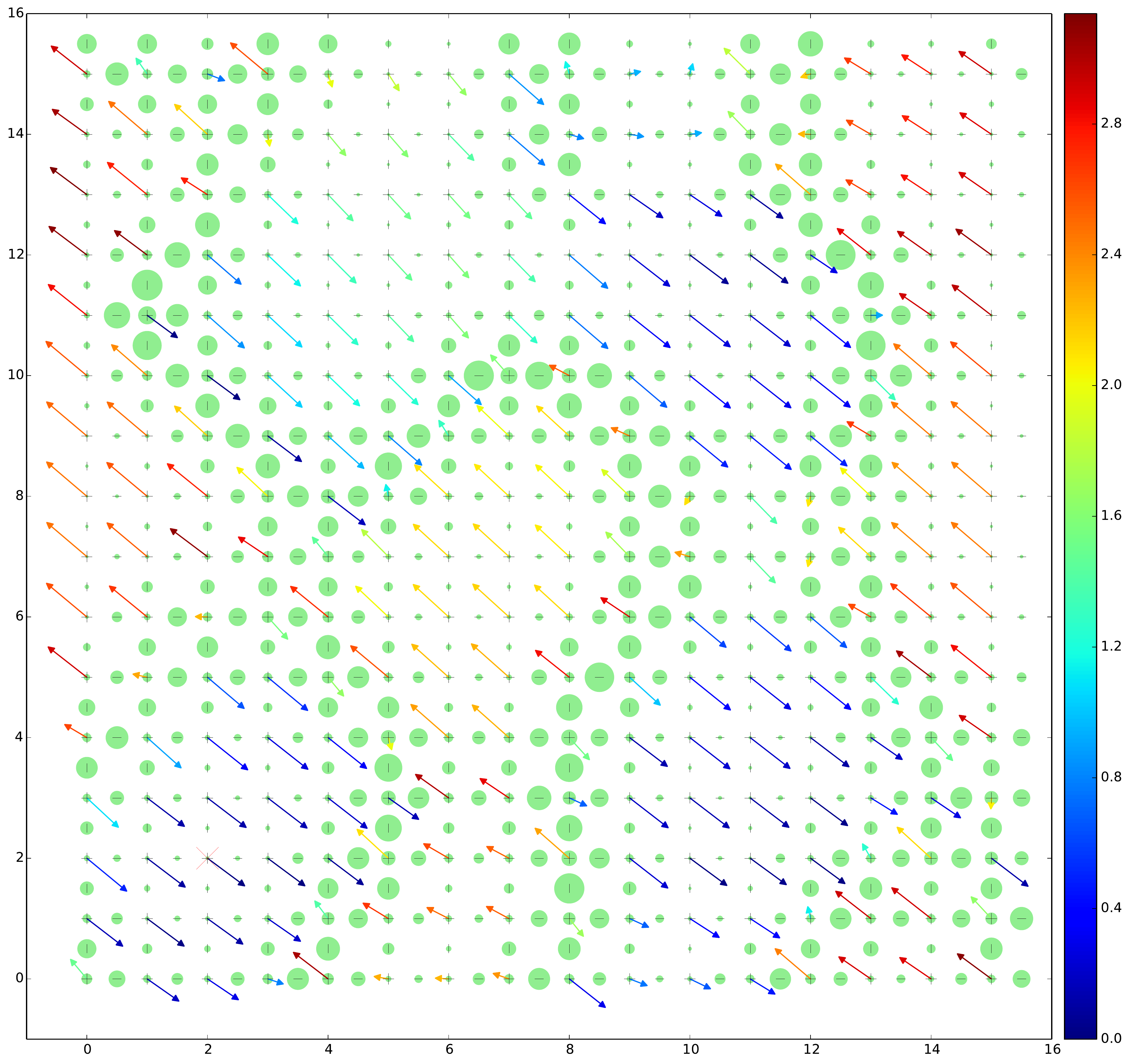}}\\
  \subfloat[2.8eV]{%
    \includegraphics[width=5.5cm]{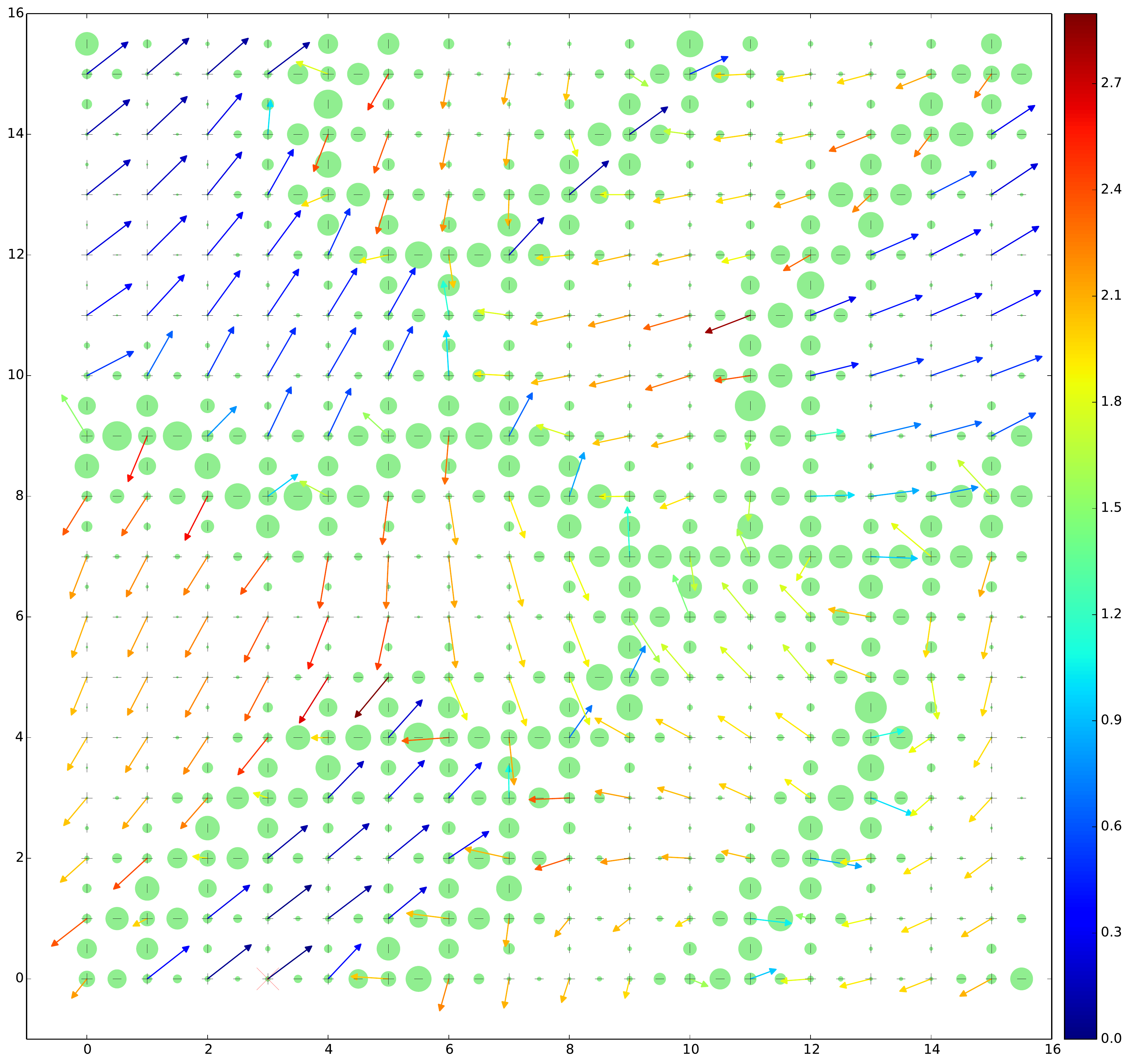}}\hfill
  \subfloat[2.9eV]{%
    \includegraphics[width=5.5cm]{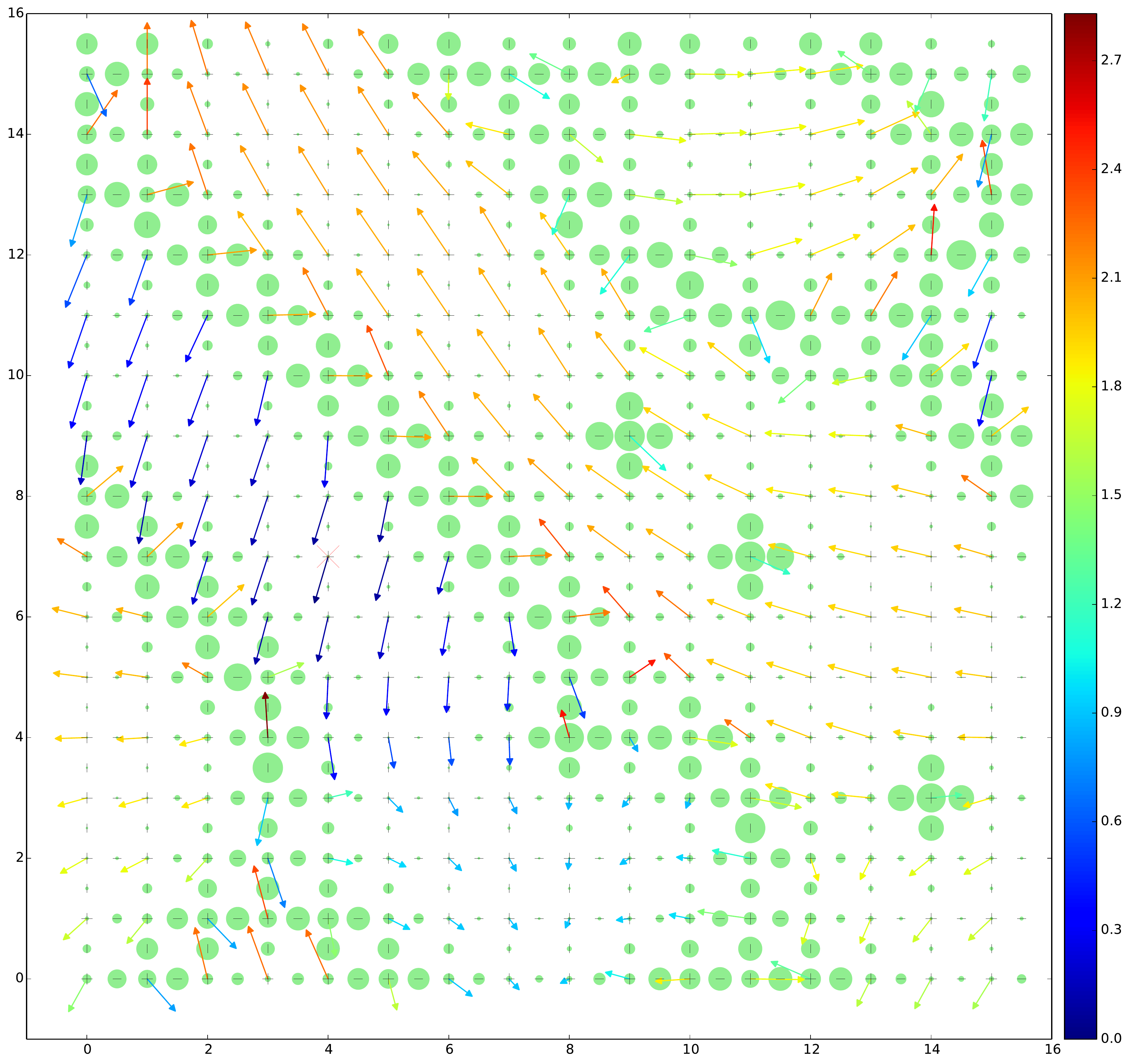}}\hfill
  \subfloat[3.0eV]{%
    \includegraphics[width=5.5cm]{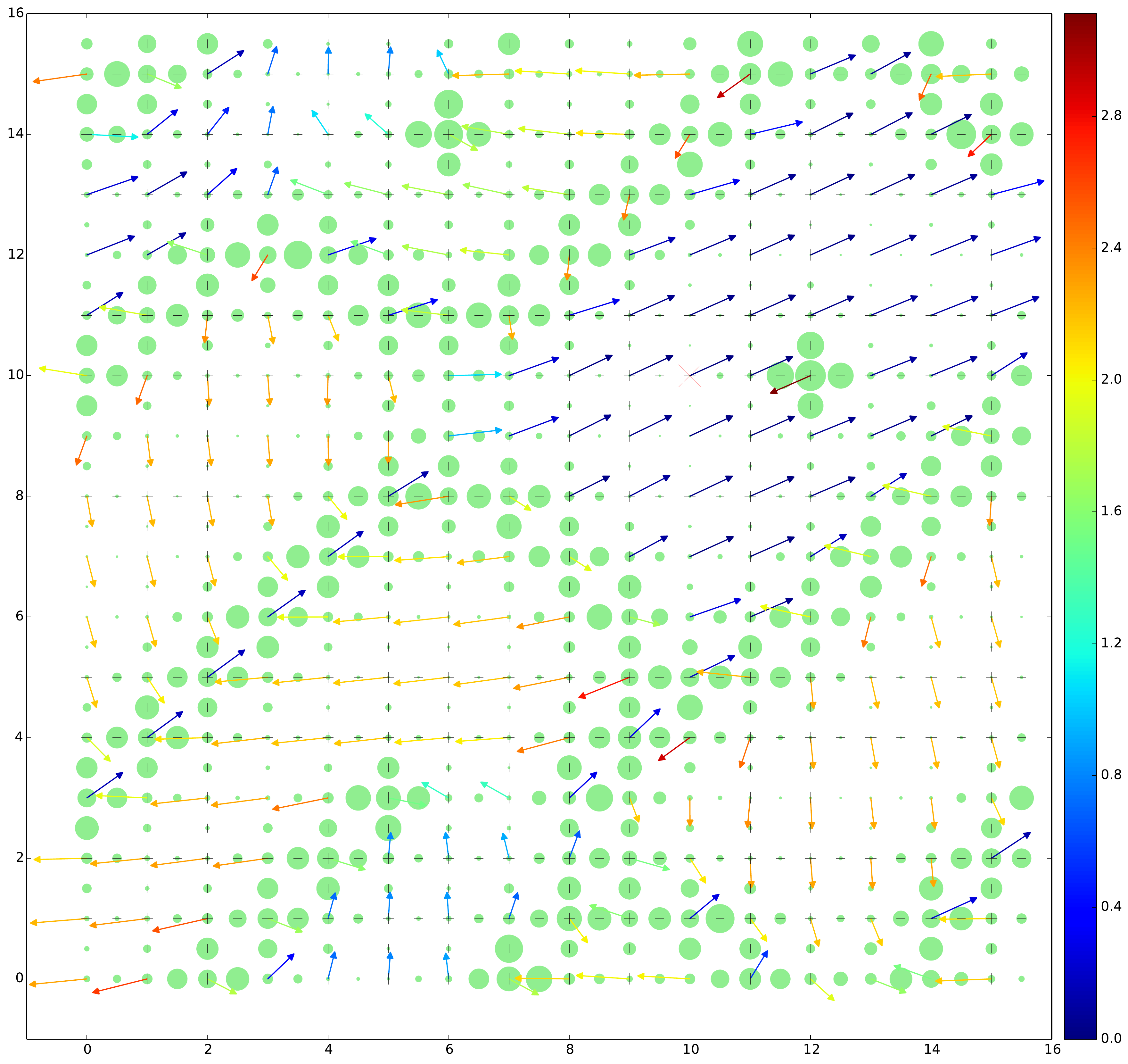}}
  \caption{
    (Color online) Evolution of the magnetic and charge order from an AFM spiral to ferromagnetic domain walls.
    A $16 \times 16$ supercell is studied for varying
    values of the charge transfer energy, $\Delta$, in the intermediate regime, at $h=1/8$.
    The total staggered spins (arrows) are plotted as a projection in the $x$-$y$ plane. The color of the arrow represents the angle between
    the spin on that site and a reference spin marked by the red ``x''. It can be thought of as a spin correlation and it runs from $(0, \pi)$.
    The spin on the O sites is negligible and omitted from the plot.
    The size of the green circles are proportional to the excess hole density from half-filling, $\delta n_{\rm excess}(\vec{r}) $.
}
  \label{fig:fine_trans_angle}
\end{figure*}

The low doped regimes for the large ($\Delta = 4.4$) and small charge transfer   ($\Delta = 1.5$) systems highlight their  differences.
In the large charge transfer regime, as discussed in Sec.~\ref{ssec:formDW},
 occupation of the O $p$-orbital is highly unfavorable compared to occupation of the Cu $d$-orbital, 
 even at very low doping.
This causes the system to find charge configurations in which the extra hole density is  concentrated and localized around some Cu sites, where a spin-flip occurs to exploit exchange energy.
In contrast, in the lower charge transfer regime, extra holes occupying an O $p$-site 
is less unfavorable. This allows more uniform charge
configurations in which the O sites carry most of the excess charge. From the uniformity, the spin order on the Cu sites is
able to largely retain AFM 
order, with smooth modulations leading to wave behavior (i.e. spin waves and spirals).

\subsection{ Intermediate Nematic Order}
\label{ssec:intermediate}

\COMMENTED{
\begin{figure*}[ptb]
\centering
  \subfloat[2.0eV]{%
    \includegraphics[width=5.5cm]{16x16_20_order.eps}}\hfill
  \subfloat[2.1eV]{%
    \includegraphics[width=5.5cm]{16x16_21_order.eps}}\hfill
  \subfloat[2.2eV]{%
    \includegraphics[width=5.5cm]{16x16_22_order.eps}}\\
  \subfloat[2.4eV]{%
    \includegraphics[width=5.5cm]{16x16_24_order.eps}}\hfill
  \subfloat[2.5eV]{%
    \includegraphics[width=5.5cm]{16x16_25_order.eps}}\hfill
  \subfloat[2.6eV]{%
    \includegraphics[width=5.5cm]{16x16_26_order.eps}}\\
  \subfloat[2.8eV]{%
    \includegraphics[width=5.5cm]{16x16_28_order.eps}}\hfill
  \subfloat[2.9eV]{%
    \includegraphics[width=5.5cm]{16x16_29_order.eps}}\hfill
  \subfloat[3.0eV]{%
    \includegraphics[width=5.5cm]{16x16_30_order.eps}}
  \caption{
    (Color online) Magnetic and charge order plot for a series of $16 \times 16$ lattices for varying
    values of the charge transfer energy in an intermediate regime between the MDW and spin spiral phases.
    The total spin (arrows) are plotted as a projection in the $x-y$ plane. The total spin, $S_{tot}(\vec{r})$,
    on each site is given by the color gradient on the right. The spin on the Oxygen p-bands is negigible and omitted from the plot.
    The size of the green circles are proportional to the excess hole density from half-filling, $\delta n_{excess}(\vec{r}) $.  
}
  \label{fig:fine_trans}
\end{figure*}
}

Within GHF, the MDW phase discussed in Sec.~\ref{ssec:magDW} is sustained at high $\Delta$ down to $\Delta\sim4.2$, while the spin spiral phase discussed in Sec.~\ref{ssec:spirals} exists up to 
$\Delta \sim 2$. 
This is for hole doping of $h=1/8$ and the parameter values given in Table~\ref{tab:param_table}. Clearly the phase boundaries in $\Delta$ as well as the nature of the phases
can vary with the other parameters.  
We next investigate charge transfer values between the two regimes, scanning $\Delta =2$-$3$. 
Our goal was to determine what happens in the intermediate region between the MDW and the spiral 
states, whether additional phase(s) exists, and how the transition occurs.

\COMMENTED{ 
what happens in the intermediate region between the MDW phase at large $\Delta$ and the spin spiral 
at low $\Delta$. 
Is there some sort of transition? Is the transition smooth or abrupt? And what do these phases look like?
}

Fig.~\ref{fig:fine_trans_angle} shows order plots for an $L_x\times L_y=16 \times 16$ supercell, i.e.~Cu$_{256}$O$_{512}$, at $h = 1/8$ for varying values
of the charge transfer energy.
Systems with $\Delta \leq 2.0$ lie in the spiral
phase, while systems with $\Delta \geq 4.3$ lie in the MDW phase.
Between the two, there appears to be a crossover 
regime.
Though the charge and spin order are 
unique at each $\Delta$ value as we scan through the transition region, there are common features.
At lower values of the charge transfer energy $\Delta$, the excess charge is distributed more uniformly
across the O $p$-orbitals in the lattice. When $\Delta$ 
is raised, the excess charge  accumulates into
predominantly diagonal lines on the lattice. Up until a sufficient value of $\Delta \sim 2.8$, these  lines
are mostly centered on the O $p$-sites. As the charge transfer energy is increased, the lines of excess charge become more
rigidly locked at $45^{\circ}$ ($135^{\circ}$) angles. 

Inspecting the spin order in conjunction with the charge order, we also see clearly the effect of doping 
on the system.
Uniform excess holes that lie on the O $p$-orbitals allow for a smooth spiral modulation of the spin. 
The system is able to retain anti-ferromagnetism to leading order with an overall modulation to accommodate the excess holes.
As the charge starts to accumulate from a uniform distribution to localized diagonal lines, it creates a need
for more dramatic changes in the spin order from that at half-filling to accommodate highly localized excess holes. Away from these diagonal lines, the charge order is close to
that of half-filling at the same $\Delta$ and therefore recovers anti-ferromagnetic order. Close to these lines of excess charge,
the spin turns sharply, creating ferromagnetic domain walls that separate spin-isolated anti-ferromagnetic domains.

The excess charge on the Cu atoms leads to a staggered spin correlation between the spin in the neighboring domain and the spin on the domain wall.
The value of this staggered spin angle correlation between the two spins seems to be related to the excess hole occupation on the Cu atom.
The greater the excess hole occupation, the greater the staggered spin angle correlation is between the domain wall and the neighboring AFM domains.
At first, the $d$-sites develop frustration, i.e., ferromagnetic links, from the AFM background, creating a phase slip similar to the spin-density 
wave or stripe phase in 
the one-band Hubbard model \cite{chia-chen-prl-10,Xu-JPCM-11,science-17}. 
As $\Delta$ further increases, the Cu sites  near the excess charge line start to develop spin-isolating ferromagnetic domain wall 
order as a precursor to the MDW 
phase. Separated by the walls of localized charge density,
the AFM order in each isolated domain can be in a completely different direction (third row), 
creating nematic orders. 
The spin correlation hits a critical point at a certain excess hole occupation on the Cu atom at which the correlation is at a maximum value of $\pi$.
For example, when at $\Delta = 3.0$ some of isolated spin
flip defects which correspond to this maximum value can be seen.
Once the charge transfer energy is high enough ($\Delta \sim 4.3$) to consistently allow greater excess hole occupation on the Cu $d$-orbitals than the
O $p$-orbitals, we enter the MDW phase, in which the different domains become phase coherent.


Our results in this regime, particularly those regarding the interface between two domains, seems consistent with the spin canting phases found by Seibold et. al,
who use an unrestricted Gutzwiller approximation on the three-band Hubbard model \cite{Seibold_canting,Seibold_nematic}.
They argue that the spin canting phase is a result of the  competition between a classical diagonal stripe phase,
characterized by the localized domain walls, and linear spin spirals, characterized by the canting of the spin order near these lines.
Though their study finds this order at very low doping of $h = 0.03$, our results are consistent with 
this interplay between the localized diagonal domain walls and the spin spirals.
Unlike this spin canting phase, our nematic phase does not exhibit coherent AFM order in the domains throughout the supercell.
Rather, the spiral behavior across different domain walls occurs in different random planes, causing the 
direction of the AFM order within the domains to isolate from each other, creating the incoherent AFM domains.

\COMMENTED{
The spin order smoothly transitions from a SDW/spiral
to the magnetic domain walls. At $\Delta = 2.1$ there is some wave behavior, evident from the staggered spin plot.
By $\Delta = 2.4$, there is almost no wave behavior in the spin. What we see are anti-phase, anti-ferromagnetic domains
inbetween the localized excess charge. This spin structure is most similar to a diagonal stripe phase. This spin order 
continues up until $\Delta = 2.9$ where some of the magnetic spin flips start to occur. At $\Delta = 3.0$, we recover the MDW 
phase.}

\COMMENTED{
Fig.~\ref{fig:transition} shows order plots for varying values of the charge transfer energy from $\Delta = 4.0$ to
$\Delta = 1.5$. What we see is an obvious transition phase around $\Delta = 2.5$ that doesn't ressemble
either the magnetic domain walls nor the spiral phase.}

\section{Summary and Discussion}
\label{sec:concl}

\COMMENTED{
x -Inclusion of p-bands has led to many new interesting phases

x Varying charge transfer energy leads to two completely different phases with some crossover

--Questions not answered:

 x	-Is stripe phase truly 1-D or actually 2-D structures?

 x	-What is the true spiral wavelength at optimal doping? 8 or 9?

 x	-Is there a third order in the crossover or is it the stripe phase competing with spiral phase?

 	-Is there actually some charge order for Delta = 1.5 or does this arise from thin lattice?

	-Low doping spiral wavelength

--Mean field gives reasonable phases that can be targeted with QMC

--With new experimental results, p-bands have significant order, mean field agrees

--Parameter space is vast, needs more scanning

--How holes bind into lines--may be something similar with normal stripe orders

--How doping has effect on spiral wavelength

--Crossover around the metal-insulator transition from QMC half-filling
}

We have presented our  study of the hole-doped CuO plane applying the generalized Hartree-Fock 
approach to the three-band Hubbard model. 
We scanned values of the charge transfer energy and different doping parameters, using
``physical'' values for the other parameters as derived for lanthanum-based cuprates. 
We find that, compared to the simplest picture of the one-band Hubbard model,
the inclusion
of the Oxygen $p$-orbitals within the three-band (Emery) model leads to new
phases with characteristics potentially of direct relevance to experimental observations 
in high-$T_c$ materials.

The charge transfer energy directly affects the hole occupation on the Cu $d$-orbitals versus 
the O $p$-orbitals.
Our study showed that, even for high values of the charge transfer energy ($\Delta = 4.4$), 
where occupation of the O $p$-orbitals is highly unfavorable, there is still non-trivial hole occupation on
the $p$-sites, highlighting the importance of the three-band model. 
Though a majority of the doped holes lie on the Cu $d$-orbitals, there is significant 
ordering on near neighbor O $p$-obitals. As the charge transfer energy is lowered, 
more holes occupy the $p$-orbitals as expected. At $\Delta = 1.5$
the charge ordering is uniform with a vast majority of the doped holes occupying the O $p$-orbitals.

While varying the charge transfer energy at doping, $h = 1/8$, we find 
three distinct phases: 
Magnetic Domain Walls, Spin Spirals, and an intermediate crossover with nematic order.
In the MDW 
phase, which occurs at large $\Delta$,
the doped holes localize on diagonal lines centered on the Cu
sites. 
An overall spin flip ($\vec{S} \rightarrow -\vec{S}$) occurs in the middle along the diagonal line,
embedded in an anti-ferromagnetic background. This forms 
diagonal domain 
walls with a thickness of three Cu sites along the $x$ (or $y$) direction, with the middle one 
bearing the spin flip and most of the excess charge.
There is no phase flip between adjacent AFM domains separated by the MDW. 
Though our lowest energy solutions with finite (but large) supercells tend to show some 
two-dimensional features, it seems likely that the GHF ground state in the thermodynamic limit 
is one-dimensional, with parallel diagonal MDWs, 
with a lateral spacing of 6 Copper sites. 

 In the spiral phase, which occurs at low $\Delta < 2.0$,
the charge order is uniform with most of the doped holes in the O $p$-orbitals.
The spins on the Cu sites have a modulated anti-ferromagnetic order. The anti-ferromagnetic order
slowly turns in a randomly chosen plane (depending on the initial variational state) as it propagates 
along $x$ or $y$-direction.
The wavelength
of the spiral, for $h=1/8$ and with the {\it{ab initio}} parameters, appears to be around 8 or 9 Cu sites.

In the intermediate $\Delta$ region, there is competition between the uniform charge order of the spiral phase 
and the localized charge order of the MDW phase. A majority of the doped holes occupy the O $p$-orbitals.
Starting from an antiferromagnetic spiral state propagating along the $y$-direction ($x$-direction), 
the 
different spiral lines at different $y$ ($x$) positions gradually develop phase differences beyond the 
perfect staggered order $(-1)^y$. 
As the charge transfer energy is increased, 
holes start to localize on diagonal lines centered on 
the O $p$-sites. The spins in the Cu $d$-orbitals near such O sites
make more drastic deviations from the AFM order. As the diagonal lines of excess charge form,
the Cu spin order near it most resembles the linear spin-density wave or stripe order seen 
in the one-band Hubbard model. The diagonal lines of charge
separate anti-phase, anti-ferromagnetic domains.

It is important to keep in mind that these are results from a mean-field GHF approach. 
More accurate treatment of correlation effects can move the phase boundaries 
with respect to parameter values (or even invalidate some of the phases). 
However, experience 
from the one-band Hubbard model indicates that HF tends to capture most of the magnetic and charge 
orders qualitatively \cite{chia-chen-prl-10,Xu-JPCM-11}. 
In fact in the one-band Hubbard model, UHF with a renormalized effective $U/t$ parameter
seems to give quantitatively quite accurate results on the magnetic and charge orders \cite{Mingpu-sc-PRB}.
Our study serves as 
a starting point for future studies, and 
reveals several important candidate phases.
For certain advanced methods such as auxiliary-field quantum Monte Carlo (AFQMC) \cite{}, our 
results also provide the necessary trial wave functions.  
Furthermore, a self-consistency 
procedure coupling our GHF calculation to AFQMC will allow an even more accurate determination
of the many-body ground state. 

It will be very 
interesting, in future studies, to investigate possible connections of the 
characteristics of magnetic and nematic orders 
to superconducting order. This will require more advanced methods, since 
no superconducting order can arise
within the GHF approach adopted here. 
An approach that generalizes it would be to introduce a 
 term with paring order in the mean-field Hamiltonian,
 and couple the calculation self-consistently to a many-body calculation 
 (e.g., AFQMC) to match spin densities and  anomalous density matrix  
(pairing order parameters),
which will allow an effective pairing interaction strength to be determined.

We thank the Simons Foundation and NSF (Grant No. DMR-1409510) for their support, 
High Performance Computing at William \& Mary for their resources and help, 
Henry Krakauer, Enrico Rossi, Mingpu Qin, Peter Rosenberg, and Kyle Eskridge for useful feedback and conversations, 
and Lucas Wagner for providing us with the parameter values from Table~\ref{tab:param_table}.
The Flatiron Institute is a division of the Simons Foundation.

\end{document}